\documentclass[a4paper,fleqn]{cas-dc}

\usepackage[authoryear]{natbib}
\usepackage{graphicx,times}
\usepackage{subcaption}
\usepackage{amssymb,amsmath}
\usepackage{aas_macros}
\usepackage{ulem}

\let\oldAA\AA
\renewcommand{\AA}{\text{\normalfont\oldAA}}
\usepackage{pdflscape}

\def\tsc#1{\csdef{#1}{\textsc{\lowercase{#1}}\xspace}}
\tsc{WGM}
\tsc{QE}

\begin{document}
\let\WriteBookmarks\relax
\def\floatpagepagefraction{1}
\def\textpagefraction{.001}

\shorttitle{Periodicity of sub-pulses in a radio pulsar}    

\shortauthors{P.Tian \& P.Zhang \& W.Yang \& W.Wang \& P.Wang}  

\title [mode = title]{Periodicity of sub-pulses in a radio pulsar}  

\tnotemark[<tnote number>] 


%

\author[1,2]{Peng-Fu~Tian}
\cormark[1]
\ead{tpengfu@whu.edu.cn}
\author[1,2]{Ping~Zhang}
\author[1,2]{Wen~Yang}
\author[1,2]{Wei~Wang}
\author[3]{Pei~Wang}




\affiliation[1]{organization={Department of Astronomy, School of Physics and Technology, Wuhan University},
            city={Wuhan},
            postcode={430072}, 
            country={China}}
            
\affiliation [2]{organization={WHU-NAOC Joint Center for Astronomy, Wuhan University},
            city={Wuhan},
           postcode={430072}, 
            country={China}}
\affiliation[3]{organization={National Astronomical Observatories, Chinese Academy of Sciences},
            city={Beijing},
            postcode={100012}, 
            country={China}}




\credit{}


\cortext[1]{Corresponding author}



\begin{abstract}
Pulsars are known to manifest complex phenomena, such as nulling, sub-pulse drifting, and periodicity. Within the purview of this investigation, we have harnessed the wavelet analysis technique to scrutinize the multifaceted periodicities and sub-pulse drifting characteristics exhibited by PSR J1926-0652, discovered by the Five-hundred-meter Aperture Spherical radio Telescope (FAST). Our analysis draws upon the rich dataset acquired from the FAST ultra-wide-bandwidth receiver (UWB), meticulously examining pulse attributes encompassing an entire pulse. It is notable that the pulse apex recurrently manifests approximately every 17.11 times the pulsar's period P, individual sub-pulses exhibit a drifting phenomenon, with a phase decrement of approximately $1.04^\circ$ over each P. Intriguingly, the central phase of each sub-pulse track gradually increments over temporal evolution. Furthermore, the relative offset distribution between successive sub-pulse tracks emanating from the trailing and leading components remains comparatively stable, with a central tendency of approximately $\sim$6.87$\pm$2.56 P. Most notably, derived from the outcomes of wavelet analysis, we ascertain a negative correlation of -0.98 between the periods of sub-pulses and their drifting rates, alongside the intrinsic period of sub-pulses identified at 28.14 seconds.
\end{abstract}




\maketitle

\section{Introduction}
Pulsars are rotating neutron stars which are highly magnetized and radiating super powerful electromagnetic radiation from theirs poles. The pulsars' electromagnetic radiation is almost visible and detectable across the entire spectrum \citep{2022ARA&A..60..495P}. Since the discovery of pulsars in 1968, The emerging research field is experiencing rapid growth and has yielded abundant research outcomes and significant advancements; however, the underlying physical mechanisms about the radio emission of pulsars remain unidentified. This is primarily attributed to the numerous unexplained physical phenomena associated with pulsars, including Fast Radio Bursts (FRB) \citep{2007Sci...318..777L}, repeaters \citep{2022Natur.609..685X}, nulling \citep{1970Natur.228...42B}, sub-pulse drifting \citep{1970Natur.228..752B,2002A&A...393..733E,2013A&A...552A..61H}, glitches \citep{2017ARA&A..55..261K,2018Natur.556..219P,2011MNRAS.411.1917W,2013MNRAS.432.3080K}, and dwarf pulses \citep{2023NatAs.tmp..177C}, and among others. 

Sub-pulse drifting is a key phenomenon for understanding the pulsar emission mechanism, in which the relative phase of the periodic sub-pulse signal shifts in organized patterns from one pulse to the next \citep{1968Natur.220..231D}. Some studies aimed to explore the nature of sub-pulse were conducted after that \citep{1969Natur.223..797T,1970Natur.228...42B,1970Natur.228..752B,1970ApJ...159L..89S,1971ApJ...167..273T,1973ApJ...182..245B,1973MNRAS.163...29P}. Generally, longitude-resolved fluctuation spectrum based on Fourier technique is mostly widely employed to learn about the properties of sub-pulses drifting \citep{1970Natur.227..692B, 1973ApJ...182..245B}. 

The underlying mechanism responsible for the drifting phenomenon is really complicated and challenging to explain, given the diverse array of pulse profile shapes and drifting patterns observed across different pulsars \citep{2004A&A...421..681E,2005MNRAS.357..859R,2006A&A...445..243W,2007A&A...469..607W,2016ApJ...833...29B,2017ApJ...836..224M,2019MNRAS.482.3757B,2023MNRAS.520.4562S}. Anyway, the phenomenon of sub-pulse drifting stands out as a highly potential process for interpreting the fundamental physical mechanisms at play. The most well known model to explain the sub-pulse drifting phenomenon is the sparking gap model, in which sub-pulses are considered as sub-beams arising from discrete stable emission regions \citep{1975ApJ...196...51R,1986ApJ...301..901R}. This model, in which these emission regions are arranged in a 'carousel' pattern rotating around the magnetic dipole axis, finds robust evidence in the observations provided by \citep{1999ApJ...524.1008D,2001MNRAS.322..438D}, and its scope of applicability has been broadened \citep{1980ApJ...235..576C,1982ApJ...263..828F,2000ApJ...541..351G,2003A&A...407..315G,2004ApJ...616L.127Q}. While this exceptional model stands as one of the most successful in the domain of emission models, it is worth noting that certain discrepancies with its predictions have also emerged \citep{1985MNRAS.215..281B,2005MNRAS.357..859R,2008MNRAS.385..606M,2009MNRAS.398.1435B}. Additionally, recent studies have indicated shortcomings in the carousel model \citep{2012ApJ...752..155V,2017MNRAS.464.2597W,2017ApJ...845...95S,2022MNRAS.514.4046W,2022ApJ...936...35B,2023ApJ...959...92B}, as well as other interpretations \citep{2020MNRAS.496..465B}.


The study of sub-pulse drifting in pulsars has not only broadened our understanding of these enigmatic objects but has also contributed significantly to our comprehension of fundamental astrophysical processes. By delving into the mysteries of sub-pulse drifting, we can decipher the underlying mechanisms that govern pulsars, shedding light on their complex nature.

In this study, we primarily investigate sub-pulse drifting and periodicity phenomena within PSR J1926-0625, a pulsar discovered by the Five-hundred-meter Aperture Spherical Radio Telescope (FAST) using its ultra-wide-bandwidth receiver (UWB) in August 2017 and subsequently confirmed by observations from the Parkes Radio Telescope in October of the same year \citep{2019ApJ...877...55Z}. 

The paper is structured as follows. In section \ref{sec:obs}, we show the observation of FAST and describe the methods that utilized for analysing. In section \ref{sec:res}, we present and analyze the main results. In section \ref{sec:dis} and \ref{sec:con}, we make some discussion and conclude the paper.

\section{Observation and Method}
\label{sec:obs}
The original observation projects of PSR J1926-0625 are monitored by both FAST and Parkes radio telescope. FAST is the largest single-dish radio telescope, which has the most sensitively capacity to detect pulsars. In the Parkes observations, the sensitivity was insufficient to detect the sing pulses from this pulsar; however, its excellent calibration capability facilitated measurements of polarization and long-term monitoring \citep{2019ApJ...877...55Z}. In the context of this study, we mainly focus on the properties of sub-pulse drifting, with a particular reliance on the results obtained from the FAST. Consequently, the subsequent discussion is closely associated with the FAST observations. The more in-depth examinations and observational details can be found in Zhang $et\,al.$ \citep{2019ApJ...877...55Z}.

FAST monitored PSR J1926-0652 for about 50 minutes on November 28, 2017, with a wide-bandwidth receiver whose bandwidth is ranging from 270 MHz to 1.6 GHz. It is noteworthy that early data from FAST were incomplete, typically consisting of only one polarization channel and data within the low-frequency band (270-800 MHz) available for analysis. The underlying reasons for this limitation remain unclear; however, it is not within the covering of our current investigation.

Zhang $et\,al.$ had recorded 1921 single pulses with a 100-microsecond time resolution, they used {\tt DSPSR} program \citep{2011PASA...28....1V,2004PASA...21..302H} to extract pulses with 512 phase bins each pulse period. The complete pulse stack can be derived from FAST's single-pulse data, which comprises only a single polarization channel and is averaged over the bandwidth ranging from 270 to 800 MHz. In Zhang $et\,al.$ \citep{2019ApJ...877...55Z}, they labeled the several episodes that occurred during this observation with Burst and numbered from 1 to 6. Here in the work of this study, we re-processed the data of the Burst 3 episode to demonstrate the feasibility and advantages of wavelet analysis, a new approach to temporal periodicity in the analysis of evolutionary complexity from small to large timescale ranges.

We have utilized wavelet analysis \citep{1992AnRFM..24..395F,1998BAMS...79...61T,2022MNRAS.513.4875C,2022MNRAS.517..182C,2023arXiv230714015T}, longitude-resolved fluctuation spectrum \citep[LRFS,][]{1970Natur.227..692B, 1973ApJ...182..245B} and two-dimensional fluctuation spectrum \citep[2DFS,][]{2002A&A...393..733E,2006A&A...445..243W} for periodic signal analysing in this work.  

Wavelet transforms are found on group theory and square integrable representations, enabling the decomposition of a signal or field into spatial, scaling, and potentially directional components. 

        




The wavelet transformation of a continuous function $f(t)$ is given by the convolution of itself with an mother wavelet $\psi (\eta)$. Meanwhile, the mother wavelet $\psi$ at each scale is normalised to have unity energy ($ \int_{-\infty}^{\infty}\psi\psi^{*}d\eta=1$, where the $^{*}$ marks the complex conjugate). for a continuous $f$:
\begin{equation}
\begin{aligned}
    W(t,s)=\int\limits_{-\infty}^{\infty}f(t^\prime)\frac{1}{\sqrt{s}}\psi^{*}\left [ \frac{(t^\prime-t)}{s} \right ] dt^\prime,
\end{aligned}
\end{equation}
where $t$ and $s$ are the time and wavelet scale respectively. The scale factor $s$ satisfies the following normalisation,
\begin{equation}
\begin{aligned}
    \int\limits_{-\infty}^{\infty}\psi\left [ \frac{(t^\prime-t)}{s}\right ]\psi^{*}\left [ \frac{(t^\prime-t)}{s}\right ]\,dt^\prime=s,
\end{aligned}
\end{equation}

However, in practical scenarios, we deal with discrete time series $x_n$, whose wavelet transformation is then given by
\begin{equation}
\begin{aligned}
    W_n(s)=\sum\limits_{n^\prime=0}^{N-1}x_{n^\prime}\sqrt{\frac{\delta_t}{s}}\psi^{*}\left [ \frac{(n^\prime-n)\delta t}{s} \right ],
\end{aligned}
\end{equation}
where $n$ is the localized time index, $N$ is the number of points in time series $x_n$ \citep{1995PhT....48g..57K}, and $\delta_t$ is time step. Typically, the selection of an appropriate mother wavelet is contingent upon the characteristics of the time series data and the expected information one seeks to extract from the signal. There are numerous mother wavelets \citep{1995PhT....48g..57K}, here in this work, for obtaining more accurate frequency information regarding the drift of sub-pulses, we employed the Morlet wavelet (m=6),
\begin{equation}
\begin{aligned}
\psi(\eta)=\pi^{-1/4}e^{im\eta}e^{-\eta^2/2}
\end{aligned}
\end{equation}
where the $m$ is wavelet parameter, which controls the number of oscillations in mother wavelet and determines the frequency and time resolution of the wavelet transform \citep{2004SoPh..222..203D}. Typically, as the value of m increases, the number of oscillations in mother wavelet increases, resulting in a higher accuracy of frequency resolution in the wavelet transform, while the time resolution appears to be lower. Conversely, when m is relatively small, the situation is reversed: time resolution becomes more accurate, but the uncertainty in frequency resolution increases \citep{herrmann2014time}.

For the others methods, we should initially start from pulse stack $S_{ij}$ (Figure \ref{fig:sweep}), which is formed by dividing a one-dimensional time series composed by consecutive sub-pulses into a two-dimensional sequence based on the pulse rotation period $P$ and chronological order. Each row of this sequence is corresponding to 360 phase degrees of rotation. Subsequently, the average intensity over pulse longitude,representing the integrated pulse profile, is expressed as:
\begin{equation}
\begin{aligned}
\mu_i=\frac{1}{N}\sum\limits_{j=0}^{N-1}S_{ij},
\end{aligned}
\end{equation}
where $i$ and $j$ are the longitude bin and pulse number respectively, so $S_{ij}$ represents the measured intensity at a specific longitude phase bin and pulse number. N is the number of pulses. Then, we can further calculate the longitude-resolved variance $\sigma_i^2$
\begin{equation}
\begin{aligned}
\sigma_i^2=\frac{1}{N}\sum\limits_{j=0}^{N-1}(S_{ij}-\mu_i)^2, 
\end{aligned}
\end{equation}
and modulation index $m_i$
\begin{equation}
\begin{aligned}
m_i=\frac{\sigma_i}{\mu_i}, 
\end{aligned}
\end{equation}
$\sigma_i$ indicates the presence of sub-pulse modulation, while the index $m_i$ is a measure of intensity variation from pulse to pulse, thus is an indicator of the existence of sub-pulses. The modulation index does not provide insights into the specific patterns of sub-pulse modulation. Therefore, the prerequisite step involves computing the LRFS to discern and characterize regular intensity variations.   

The Two-dimensional Fourier transform is primarily employed to decompose the input function into a summation of complex exponentials, which is considered as an extension of the method of LRFS \citep{2002A&A...393..733E}. The definition of the two-dimensional Fourier transform of a function $f(x,y)$ is as follows:
\begin{equation}
\begin{aligned}
g(u,v)=\mathcal{F}[f(x,y)]=\int_{-\infty}^{\infty}\int_{-\infty}^{\infty}f(x,y)e^{-2\pi j(ux+vy)}dxdy,
\end{aligned}
\end{equation}
$\mathcal{F}$ is the Fourier transform. The inverse of the two-dimensional Fourier transform is:
\begin{equation}
\begin{aligned}
f(x,y)=\mathcal{F}^{-1}[g(u,v)]=\int_{-\infty}^{\infty}\int_{-\infty}^{\infty}g(u,v)e^{2\pi j(ux+vy)}dudv,
\end{aligned}
\end{equation}
generally, $x$ and $y$ are spatial and temporal, while $u$ and $v$ are spatial and temporal frequencies. For the complex function $g(u,v)$, it can be represented as $g(u,v)=g_R(u,v)+jg_I(u,v)$, where $|g(u,v)|$ signifies the magnitude spectrum, while $arctan(g_I/g_R)$ denotes the phase angle spectrum. 

For the discrete case data, the Fourier transform is two-dimensional Discrete Fourier Transform (2-D DFT), which is

\begin{equation}
\begin{aligned}
g(u,v)=\sum\limits_{j=0}^{N_x-1}\sum\limits_{k=0}^{N_y-1}f(j\Delta x, k\Delta y)e^{-2\pi i(uj\Delta x+vk\Delta y)},
\end{aligned}
\label{eq:d2DFS}
\end{equation}
The ranges for $u$ and $v$ are defined as $-1/2\Delta x< u<1/2<\Delta x$ and $-1/2\Delta y< v<1/2<\Delta y$, where $\Delta$ represents the data sampling intervals along the $x$ and $y$ axes, respectively. This criterion is commonly referred to as the Nyquist-Shannon sampling theorem \citep{1928TAIEE..47..617N,1949IEEEP..37...10S}. It should be noted that the function $f$ in equation \ref{eq:d2DFS} is discretely and uniformly sampled, which is the pulse stack $S_{ij}$ in this work.

2-D DFT is applied to the pulse stack $S_{ij}$, the result of which is the Two-Dimensional Fluctuation Spectrum (2DFS) \citep{2002A&A...393..733E,2006A&A...445..243W}.



\section{Results}
\label{sec:res}

\begin{figure}[ht]
    \centering
    \includegraphics[width=1\linewidth]{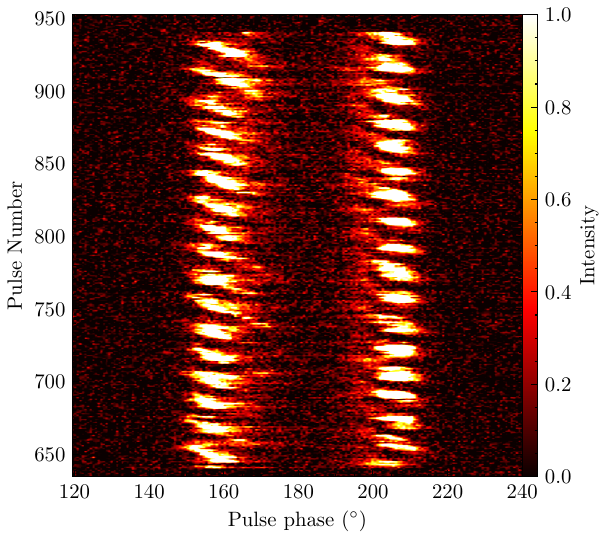}
    \caption{Pulse stack of Burst 3 (pulse number versus pulse phase) as indicated in \citep{2019ApJ...877...55Z}. Data are averaged from FAST observable band from 270 to 800 MHz of single polarization channel. Colorbar indicates the normalized intensity.}
    \label{fig:sweep}
\end{figure}

In comparison to other episodes, Burst 3 exhibits a notably greater completeness, persisting for a total duration of approximately 300 pulse periods. Moreover, during this interval, the evolution patterns of the two peaks of the pulses with respect to pulse phase (referred to as $P_2$) and pulse number (referred to as $P_3$) are highly discernible, which significantly facilitates our research endeavors. Figure \ref{fig:sweep} shows the pulse stack of the Burst 3, the leading and trailing components are both exhibiting periodicity and sub-pulse drifting. While the drifting pattern appears to follow a clearly discernible linear variation between pulse phase and pulse number. Meanwhile, it is discernible from Figure \ref{fig:sweep} that as the phase number ascends, the central phase of the leading component gradually rises. Conversely, the central phase of the trailing component appears to remain constant, yet in actuality, it undergoes a exceedingly subtle decreasing. 

\begin{figure*}
    \centering
    \includegraphics[width=0.8\linewidth]{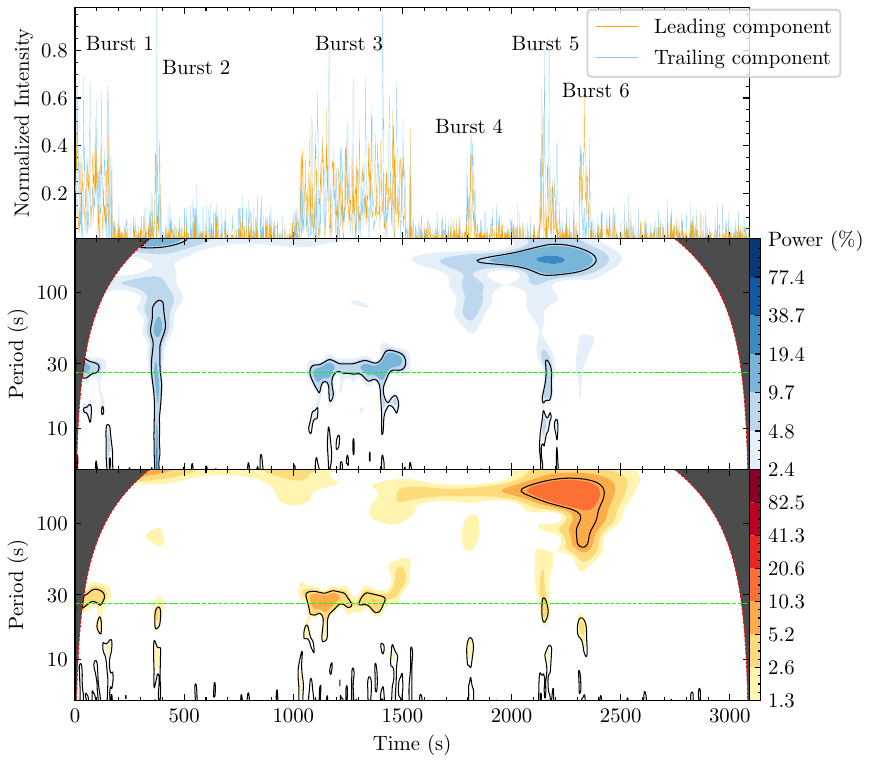}
    \caption{Wavelet analysis results of leading and trailing components. Top: the time series of two components along the pulse number. Middle and bottom: the wavelet analysis results of time series of leading and trailing components. The dotted horizontal auxiliary lines in the middle and bottom panels emphasize the position of the central peak of the sub-pulses' period over time.}
    \label{fig:wavelet}
\end{figure*}


Subsequently, we conducted a wavelet analysis on the temporal sequences of the leading and trailing components of all six episodes mentioned in the study by Zhang $et\,al.$, as depicted in Figure \ref{fig:wavelet}. The results elucidate the presence of a periodic signal with a duration of approximately 27 seconds throughout the observational period. Notably, this signal synchronizes with the appearance of each episode, indicating its concurrent existence within every episode.

\begin{figure*}
\begin{minipage}{0.65\textwidth}
\includegraphics[width=1\textwidth]{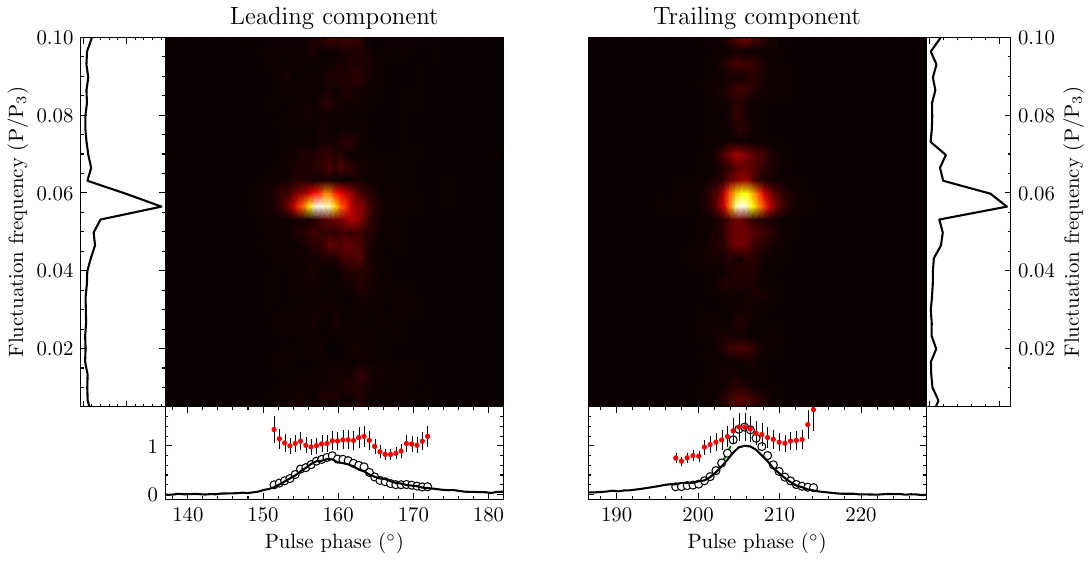}
\end{minipage}
\begin{minipage}{0.34\textwidth}
\includegraphics[width=1\textwidth]{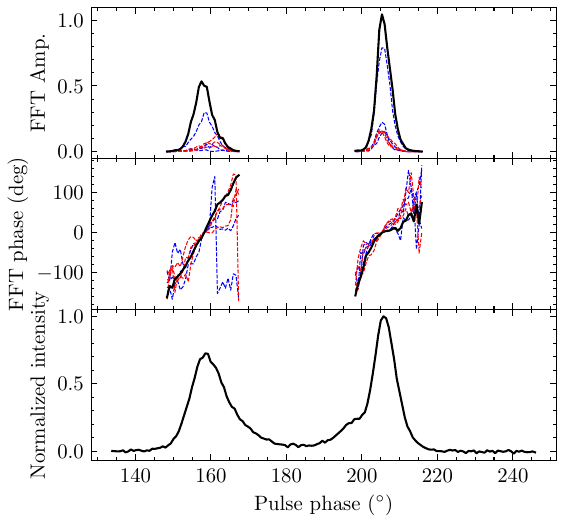}
\end{minipage}
\caption{Left: Longitude-resolved fluctuation power spectra for the leading and trailing components in the Burst 3 episode (two colored center panels). The left and right vertical side-panels show the power density spectra integrated horizontally from LRFS. The bottom panels show the integrated pulse profiles of leading and trailing components (solid black lines), the longitude-resolved modulation index (red points with error bars) and the longitude-resolved standard deviation (open circles with green dashed lines). Right: The variation of the peak frequency across the pulse longitude. Amplitude (top), phase variation (middle) and the pulse profile (bottom). The blue and red dashed curves in the top and middle panels represent regions within 3 $\sigma$ confidence interval, indicating peaks in power density spectrum whose frequencies larger and smaller than the peak frequency, respectively.}
\label{fig:LRFS_n}
\end{figure*}

On the other hand, it is interesting to observe that between the ending of Burst 3 and Burst 6, roughly spanning from 1500 to 2500 seconds, the wavelet analysis suggests the presence of a periodic signal with a period of approximately 200 seconds. This periodicity has much longer period than the drifting periodicity, exhibiting a pronounced similarity to the characteristics of periodic nulling \citep{2017ApJ...846..109B}. This is highly conceivable, as the nulling has complicated underlying physical mechanism and appears to be non-random, moreover, periodic nulling and periodic amplitude modulations are separate types of emission features unrelated to sub-pulse drifting \citep{2017ApJ...846..109B,2020ApJ...889..133B}. 
Meanwhile, we have noticed that the duration of nulling occurring between the 6 bursts is not consistently constant, which implies that the observed periodicity can also be attributed to the quasi-periodic nulling phenomenon \citep{2006Sci...312..549K}. However, due to the limited amount of observational data, we cannot conclusively determine whether this periodicity stems from periodic nulling or it may be considered as a mere stochastic event.

\begin{figure}[h]
    \centering
    \includegraphics[width=1\linewidth]{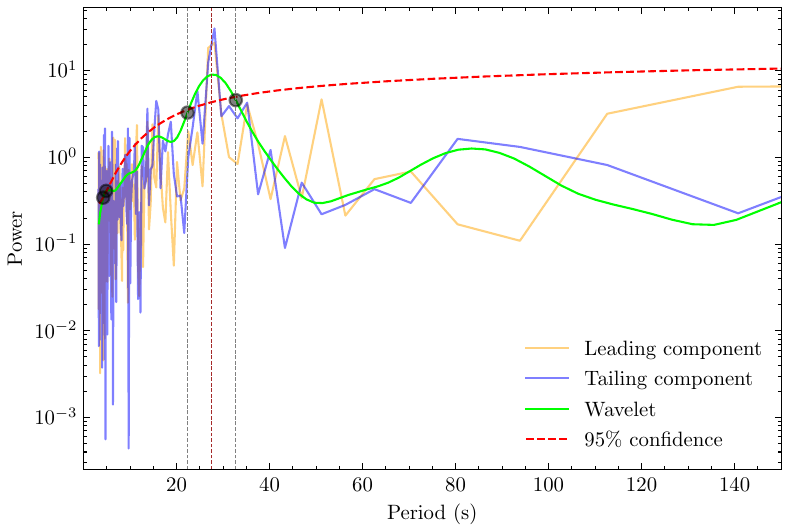}
    \caption{The PDS of leading and trailing components of pulse. The solid lines colored with orange and purple are Fast Fourier Transformation (FFT) results of leading and trailing components respectively. The lime solid line and red dashed lines represent wavelet result and 95$\%$ significance. The brown and gray dotted vertical lines are indicating the central and width of the peak (27.53$_{-5.17}^{+5.21}$ s).}
    \label{fig:pds1}
\end{figure}

\begin{figure}[hb]
    \centering
    \includegraphics[width=1\linewidth]{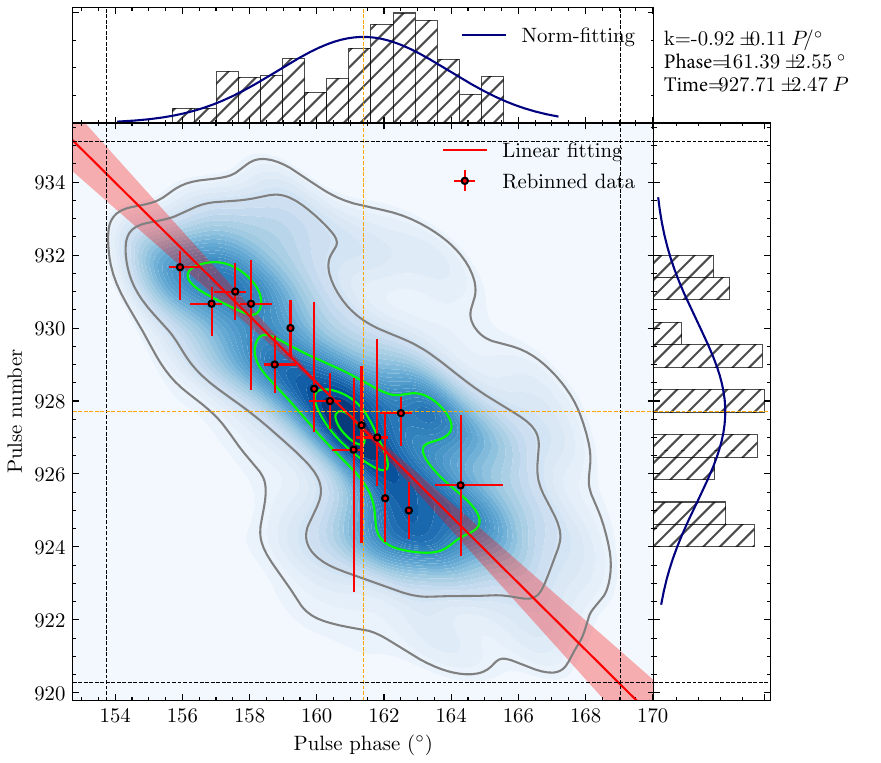}
    \caption{Contour plot of one of the sub-pulses in leading component. The red solid linear line is the linear fitting line. The contours indicate the sub-pulse track, while the colorbar means the intensity of the pulse. Scattering points are the rebinned data of the sub-pulse. Phase and Time are the coordinates of the pulse in phase and pulse number. k is the slope of the linear fitting result.}
    \label{fig:Boxhead}
\end{figure}

Regard to periodic signal in Figure \ref{fig:wavelet}, we deduce its periodicity (P$_3$) as 27.53$_{-5.17}^{+5.21}$ seconds through the power density spectrum (PDS) analysis employing wavelet analysis (also see Figure \ref{fig:pds1}). Furthermore, the findings in Zhang $et\,al.$ corroborate that both the leading and trailing components exhibit a cyclic pattern with an average period of 17.33P (approximately 27.88 seconds), aligning harmoniously with our own outcomes. In the research conducted by Zhang $et\,al.$, there are some pulse phases where the error bars of P$_3$ exhibit significant magnitudes, presenting a apparent disparity when compared to our findings. This discrepancy appears to stem from Zhang $et\,al.$'s consideration of data points situated at the periphery of the pulse profiles, where clarity is somewhat lacking, thereby resulting in substantial errors. It is conceivable that the deviation of 17.33 P, as derived by Zhang $et\,al.$, from our own results is consequently attributed to this factor. Meanwhile, Figure \ref{fig:LRFS_n} exhibits the longitude-resolved fluctuation power spectra of leading and trailing components in Burst 3, in which the two light spots in the heatmaps prominently exhibit a clearly delineated periodic characteristic, accompanied by an elevated power level within the associated pulse phase interval. The $P_3$ values calculated from Figure \ref{fig:LRFS_n} are 17.70$_{-0.87}^{+0.59}$ P for leading component and 17.70$_{-1.18}^{+0.49}$ P for trailing component, which are slightly higher compared to the results from the wavelet analysis. Nevertheless, given the consideration of error margins, they are essentially consistent. We have also calculated  the modulation indices for pulse components, which is shown in bottom panels of Figure \ref{fig:LRFS_n}. The modulation index of the trailing component is higher than that of the leading component, suggesting that the trailing component may be more strongly modulated and emanates from a distinct sub-beam compared to the one associated with the leading component. The top and middle panels of right-hand plot of Figure \ref{fig:LRFS_n} are showing the maximum amplitude and phase variations \citep{2001MNRAS.322..438D,2019MNRAS.482.3757B} corresponding to $f_c$ for LRFS (The $f_c$ is calculated from horizontally integrated LRFS, and the corresponding error $\sigma_f$ is FWHM/2.355). The phase corresponding to the peak longitude of the pulse is set to 0, so the result is the relative phase variation. In general, the phase exhibits highly stable variations during both the leading and trailing components. However, the patterns of phase variations in the leading and trailing component are quite different. The leading component shows a almost linear phase variation, whereas the phase variation of the trailing component exhibits two behaviours. Initially, during the onset of the trailing component, the phase variation undergoes a steep slope, but it subsequently transitions to a flatter slope in the later stages. This aligns with our expectations, as illustrated in Figure \ref{fig:p3folding}. The drifting of the leading component remains nearly constant, while the drifting of the trailing component initially occurs at a relatively rapid rate. However, in the later part, the drifting pattern of the trailing component starts to become ambiguous, consequently leading to a flatter phase variation.

While exploring the periodicity among sub-pulses in our investigation, we also contemplate the nature of sub-pulse drifting. To illustrate our analysis of sub-pulse drifting, we take an single instance sub-pulse from the leading component as a case study. Initially, we regard the pulse profile of a sub-pulse as a distribution of points, with the pulse's intensity, that is, the height of contours, considered as the point's kernel density. This approach allows us to obtain the distribution of points in both the pulse phase and pulse number. In Figure \ref{fig:Boxhead}, we fit these two these two distributions with Gaussian, yielding the location of an individual sub-pulse track in both phase and pulse number dimensions. Simultaneously, we subject the scatter plot to linear regression analysis and linear fitting, consequently obtaining the Pearson correlation coefficient (PCC) and the slope of the fitting result. It is noteworthy that the patterns of sub-pulse drifting are multifarious, as evidenced by Zhang $et \,al$. \citep{2002A&A...393..733E,2005MNRAS.356...59E,2019ApJ...877...55Z}. The manifestation in the case of PSR J1926-0652 predominantly assumes a linear character. Therefore, within the scope of our study, the prerequisite for fitting the drifting sub-pulses hinges upon the assumption of their linear variation.

\begin{figure*}[hb]
    \centering
    \includegraphics[width=0.8\linewidth]{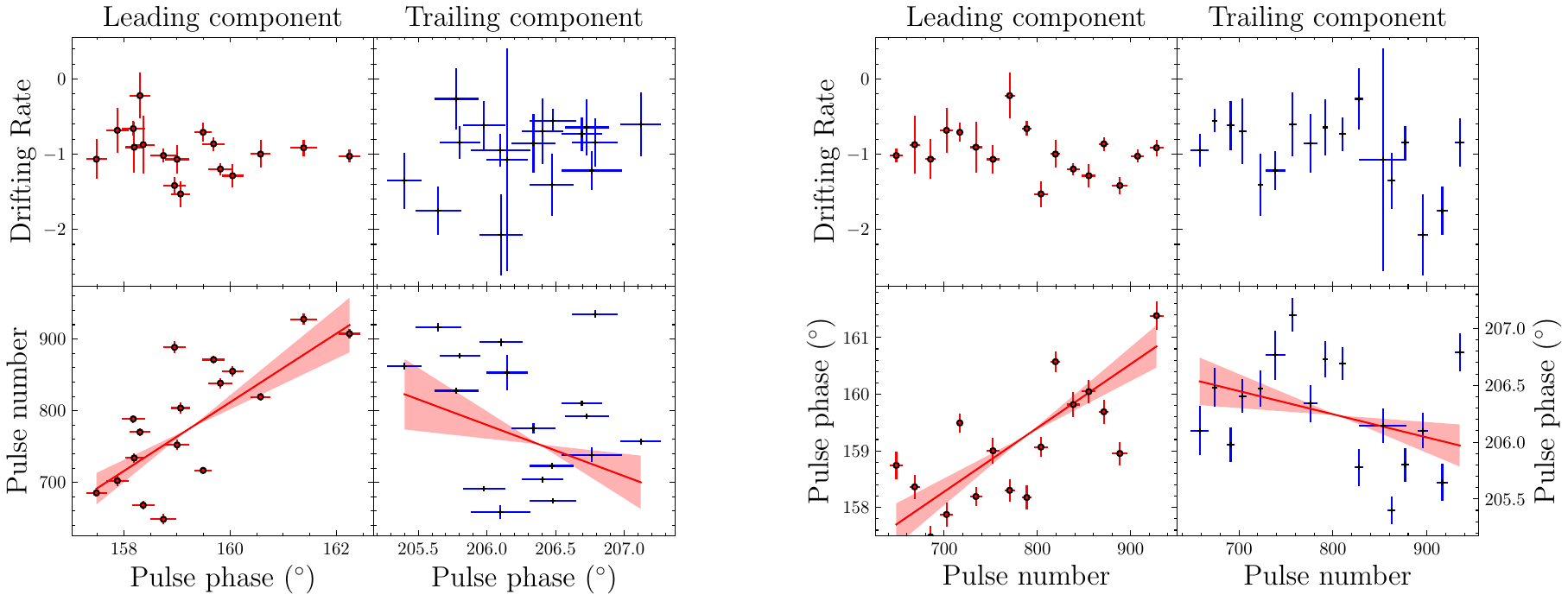}
    \caption{Top panels: drifting rates vary with central pulse number (left) and phase (right) of sub-pulse tracks. Bottom panel: central pulse numbers vary with central phases of sub-pulse tracks. The red solid lines and red regions are linear fitting and the corresponding 1$\sigma$ confidence level.}
    \label{fig:params}
\end{figure*}

In Figure \ref{fig:sweep}, we have demonstrated the pulse stack of Burst 3, where each sub-pulse manifests with clarity and distinctiveness. This really helps our subsequent analysis. For each individual sub-pulse, we manually demarcated rectangular regions in Figure \ref{fig:sweep}, akin to the result depicted in Figure \ref{fig:Boxhead}. Consequently, for every sub-pulse, whether belonging to the leading component or the trailing component, the analysis in Figure \ref{fig:Boxhead} yielded a series of parameters, including $k$ (also known as drifting rate), $Phase$, and $Time$. For these parameters, we examined their respective distributions with respect to pulse phase and pulse number, as illustrated in Figure \ref{fig:params}. 

\begin{figure}[h]
    \centering
    \includegraphics[width=1.\linewidth]{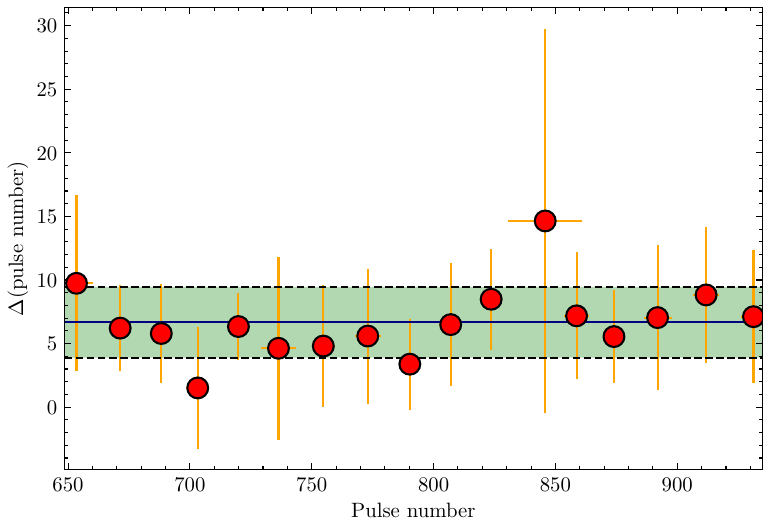}
    \caption{The distribution of offsets between peaks of the emission from the trailing and leading components as a function of the pulse number. The value of y-axis indicates the difference between the pulse numbers of tailing and leading components. The navy solid line shows the mean value of 6.87P, and the dashed black lines represent the 1-$\sigma$ confidence interval.}
    \label{fig:deltanumber}
\end{figure}

\begin{figure*}[h]
    \centering
    \includegraphics[width=0.8\linewidth]{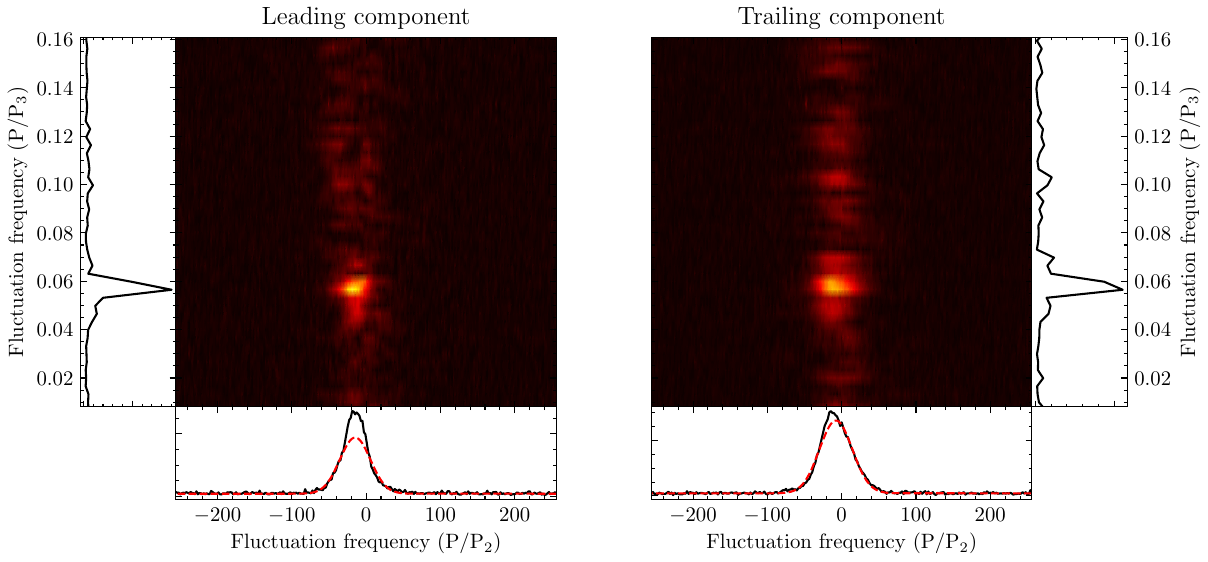}
    \caption{The 2DFS of leading and trailing components. The bottom and side panels show the power integrated horizontally and vertically from the 2DFS respectively. The red dashed curves in the bottom panels are the best gaussian fitting of the vertically integrated profiles.}
    \label{fig:secondp}
\end{figure*}

The drifting rates exhibit almost no linear variation with central phase and pulse number of sub-pulse tracks, averaging around -0.97 and -0.96 for the leading and trailing components, respectively. However, whether the variation in drifting rate occurs with respect to central phase and pulse number, it appears to exhibit a linear trend. This property, is similar to the fluctuations of period over time observed in wavelet analysis, motivates our subsequent analysis of the relationship between the periods and drifting rates of sub-pulses.

Another intriguing finding is that the pulse stack results in Figure \ref{fig:sweep} reveal some evolution among the 35 sub-pulse tracks (comprising 17 in the leading component and 18 in the trailing component) within Burst 3. Examining the distribution results between the central phases and pulse numbers of sub-pulse tracks reveals a consistent evolution in the leading component and potentially in the trailing component. In the leading component, the phase of sub-pulse tracks evolves by approximately $47.90\pm12.58$ ($P/^\circ$), indicating a progression of 47.90 pulse cycles for a one-degree approach to the central phase of the pulse stack. However, the relationship between phases and pulse numbers for the sub-pulse tracks in the trailing component is more ambiguous, with a substantially greater value of $71.32\pm50.14$ ($P/^\circ$). This suggests an almost 71.32 pulse cycle requirement for sub-pulse tracks to approach the central phase by one degree. It is noteworthy that the results for the trailing component may entail a significant uncertainty of error bars. However, it can be affirmed with certainty that the total phase shift of sub-pulse tracks in the trailing component is undeniably lesser than that in the leading component.

Remarkably, the bottom panels of Figure \ref{fig:params} and the findings in Figure \ref{fig:sweep} reveal an intriguing phenomenon. Drifting sub-pulses in the leading and trailing components exhibit an alternating pattern, as illustrated in Figure \ref{fig:deltanumber}, which shows no dependence on frequency \citep{2009MNRAS.398.1435B}. For further analysis, we have conducted a statistical analysis of the temporal intervals (as a function of pulse number) between adjacent sub-pulses within the trailing and leading components in Figure \ref{fig:deltanumber}. To obtain these results, we first extracted the central phase and pulse number for each sub-pulse track from Figure \ref{fig:Boxhead}. Subsequent to this, the obtained results were systematically arranged in accordance with pulse number, and then the temporal intervals between consecutive trailing and leading components were calculated. The outcomes reveal that all data points exhibit values greater than zero, signifying the alternatively appearing of sub-pulses in both leading and trailing components. Furthermore, the results suggest a semblance to a sinusoidal pattern; however, due to the lack of data points, we are unable to draw such a conclusion. This offset between leading and trailing components is most possible attributed from the geometry, and can be related with number of sparks, drifting rates and line of sight \citep{2009MNRAS.398.1435B}.



Finally, in Figure \ref{fig:secondp}, we computed the 2DFS of Burst 3, aiming to elucidate the nature of sub-pulse drifting from an alternative perspective. The 2DFS is developed for the investigation of periodic phenomena within the pulse stack. The fundamental operation entails the decomposition of two-dimensional data into plane waves. This process is frequently encountered in the analysis of the dynamic spectra of pulsar data, where it is often obscured by interference from foreground interstellar medium scintillation \citep{2001ApJ...549L..97S,2005MNRAS.358..270W,2010ApJ...717.1206C}. As illustrated in Figure \ref{fig:secondp}, a wide vertical bright stripe in the central region is observed, indicating the steady pulsar emission component of Burst 3's pulse stack. Apart from this, the 2DFS is central symmetry ($I(\nu_l,\nu_t)=I(-\nu_l,-\nu_t)*$, where $\nu_l$ and $\nu_t$ are the abscissa
and ordinate respectively \citep{2002A&A...393..733E}. in our findings, the segments with ordinate $\nu_t$ less than 0 are not shown). The predominant structure indicates the period modulation of the sub-pulse. The coordinates of the central position of this sub-pulse modulation component correspond to the period of the sub-pulses along pulse phase $P_2$ and pulse number $P_3$. The results shows that $P_2=24_{-1}^{+1}\,^\circ$ and $P_3=17.70_{-0.89}^{+0.56}\,P$ for the leading component and $P_2=46_{-2}^{+3}\,^\circ$ and $P_3=17.70_{-1.21}^{+0.52}\,P$ for the trailing component.


\section{Discussions}
\label{sec:dis}

In our current endeavor, we have primarily utilized observational data pertaining to Burst 3 episode from FAST's observations of the pulsar PSR J1926-0652. We have investigated sub-pulse drifting and its periodic behavior using methods including wavelet analysis, linear regression, LRFS and 2DFS. Wavelet analysis results unveil the persistent occurrence of a periodic modulation with a period of approximately 27 seconds in each pulse episode after nulling cessation. Within the wavelet analysis findings, the period manifests a measure of stability, with its width undergoes diverse degrees of evolution across different burst episodes, alternating between narrowing and widening. It is certain that, regardless of any evolutionary changes, the peak consistently maintains a duration of 27.53 seconds. Referring to Figure 6 in the study by Zhang $et\,al.$, it is possible to speculate that variations in periodic signal width may be linked to the presence or absence of faint inner components, such as C2 and C3. Simultaneously, these subtle structures are intricately associated with the phenomenon of sub-pulse drifting, thereby providing additional evidence for the correlation between the periodicity and the drifting rates of sub-pulses. It seems that the longer the burst episode lasts, the more stable its pulses and drifting rates become \citep{1975ApJ...198..661H,2019ApJ...877...55Z}. This hypothesis aligns well with the pulsar profile shape model \citep{2007MNRAS.380.1678K}, wherein profiles with varying component intensities can be attributed to specific line-of-sight intersections and component superimposition. Furthermore, Zhang $et\,al.$ offers valuable insights into the polarization angle variations between the two prominent components, which closely coincide with the findings of the rotating vector model (RVM) \citep{1969Natur.221..443R,2020A&A...641A.166P}. Given the restricted access to more extensive observational data, we abstain from delving into an exhaustive discussion regarding the physical model of the pulsar within this context.


\begin{figure}[htbp]
    \centering
    \includegraphics[width=1.\linewidth]{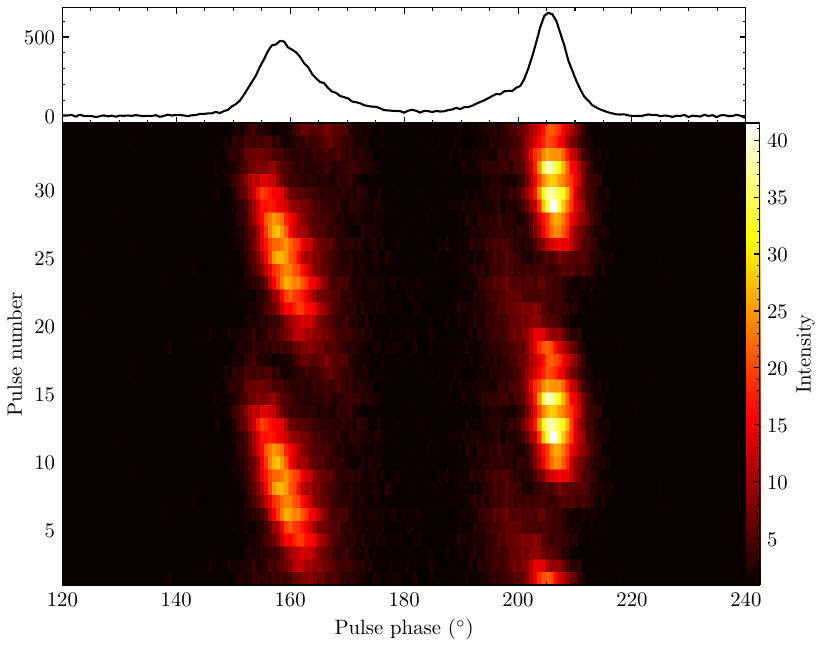}
    \caption{The $P_3$ folding result of the pulse stack in Figure \ref{fig:sweep}. The folding period is $P_3=17.49 P$, which is the intrinsic $P_3$ period deduced from MCMC result.}
    \label{fig:p3folding}
\end{figure}

On the other hand, in Figure \ref{fig:p3folding}, we have conducted $P_3$ folding on the pulse stack illustrated in Figure \ref{fig:sweep}, with the folding period determined as 17.49P based on the MCMC result. The outcomes of $P_3$ folding reveal that both C2 and C3 components significantly impact the computation of $P_3$ and drifting rate, aligning seamlessly with our preceding discussions. We have also applied the estimation depicted in Figure \ref{fig:Boxhead} to the results of $P_3$ folding, obtaining drifting rates for the leading and trailing components as $-1.43_{-0.12}^{+0.12}$ and $-1.99_{-0.59}^{+0.59}$ (see Figure \ref{fig:contourplots}), respectively. Further calculations based on the relationship, in which the drifting rate equals to $P_2/P_3$ yielded $P_2$ values of $25_{-2}^{+3}$ degree for the leading component and $35_{-10}^{+11}$ degree for the trailing component. The considerable margin of error in these results, diverging notably from the results from 2DFS in the context and Zhang $et\,al.$'s findings, is primarily attributed to the less conspicuous nature of $P_2$ in the case of PSR J1926-0652.

These results are undoubtedly related to the interior components C2 and C3. After all, these two components would also contribute weight to the statistical results. In short burst episodes, such as burst 4, 5a, and 6, the variations of drifting rate are extremely pronounced, with unambiguous drifting profiles and even positive drifting rates observed. This phenomenon is likely related to the assumption mentioned earlier, in other words, in longer burst episodes, details of sub-pulse become clear, similar to the mechanism of camera exposure, or it may just be associated with the nulling mechanism. Pulse emission and nulling phenomena in most pulsars occur simultaneously over a broad frequency range, with only a small fraction of pulsars being exceptions. However, the majority of these exceptions occur during transition phase \citep{1970Natur.228...42B,2007A&A...462..257B,2014ApJ...797...18G}. This implies that different components at different frequencies, such as C2 and C3, may not immediately appear in the onset of the burst episode after the ending of the nulling phase, leading to incomplete sub-pulse profiles and potential errorbars in the regression analysis of drifting.

As for the drift in the central phase of the pulse stack, the carousel model provides a reasonable explanation, considering the emission geometries as suggested by \citep{2009MNRAS.398.1435B}, however, the non-linear drift behaviour in the trailing component suggests deviation from the carousel model. The relevant parameters, magnetic inclination angle ($\alpha$), and impact parameter ($\beta$) are detailed in Zhang $et\,al.$ \citep{2006MNRAS.368.1856J}.

Most important of all, performing wavelet analysis on the averaged time series enables the determination of modulation periods at specific time points. This subsequently serves as a probing method to elucidate the relationship between drifting periodicity and the drifting rate (Figure \ref{fig:Slope-F-fitting.pdf} and Figure \ref{fig:Periodic-slope}). The findings suggest a pronounced negative correlation of -0.98 between the period and drifting rate. Concurrently, the intercept result $b_0$ depicted in Figure \ref{fig:Periodic-slope}, with a value of 28.14$^{+1.84}_{-1.86}$, indicate that the central region of the set of sub-pulse tracks forming a drift track might possess an intrinsic period of 28.14 s.

\section{Conclusion}
\label{sec:con}

In Zhang $et\,al.$, they first detected the pulsar PSR J1926-0652 using FAST's ultra-wideband receiver, aided by Parkes observations. They obtained its integrated pulse profile and observed various physical phenomena, such as nulling and sub-pulse drifting. Our work primarily focuses on the methodology and does not delve extensively into theoretical analysis. Our main findings are as follows:
\begin{itemize}
    \item[1.] Whenever 6 pulse stacks appear, a periodic modulation of approximately 27.53$_{-5.17}^{+5.21}$ seconds simultaneously emerges. The long-term periodic behavior around 200 seconds observed in the wavelet analysis is attributed to intervals between multiple burst episodes. Due to the lack of substantial observational evidence, its authenticity cannot be definitively confirmed.
    \item[2.] We manually selected sub-pulses, encompassing both outer and inner regions, and statistically analyzed their distribution by treating intensity values as two-dimensional distribution kernel densities. Linear regression and fitting were employed to determine the drifting rate.
    \item[3.] We report the separation between adjacent sub-pulse tracks within trailing and leading components, approximately $\sim$6.87$\pm$2.56 P.
    \item[4.] We computed the LRFS and 2DFS for the Burst 3 pulse stack to learn about the periodic modulations of sub-pulses.
    \item[5.] The outcomes derived from wavelet analysis have significantly aided in elucidating the periodicity and drifting rate of individual sub-pulses. Furthermore, we have found a negative correlation with a slope of -0.98 between the periodicity and drifting rate by a linear fitting, and an intrinsic period 28.14 s of sub-pulses. The comprehensive details are shown in Figure \ref{fig:Slope-F-fitting.pdf}, while a more exhaustive analysis is available in Appendix \ref{sec:appendix}.
\end{itemize}

\begin{figure}[ht]
    \centering
    \includegraphics[width=1.\linewidth]{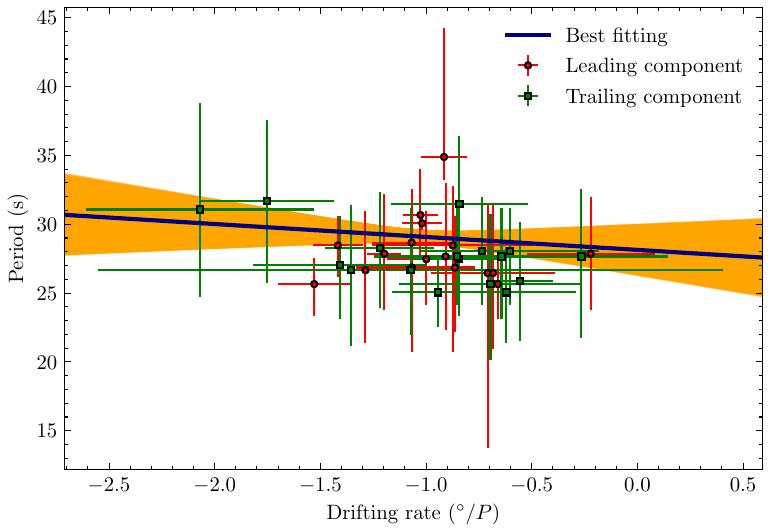}
    \caption{Period as a function of drifting rate of sub-pulses. The blue solid line indicates the best linear fitting result from Markov chain Monte Carlo (MCMC) method \citep{2013PASP..125..306F}. The results pertaining to leading and trailing components are delineated by colors green and red respectively.}
    \label{fig:Slope-F-fitting.pdf}
\end{figure}


In our previous works \citep{2022MNRAS.517..182C,2022MNRAS.513.4875C,2023arXiv230714015T,2023JHEAp..39...43T}, when dealing with one-dimensional signals, it has been customary to employ Fast Fourier Transformation (FFT) and dynamic Power Density Spectrum (dPDS). However, in addition to these approaches, we have also utilized the method of wavelet analysis for signal processing. This alternative method often yields significantly improved results, offering a more profound clarity of signal continuity and its temporal evolution characteristics. This naturally inspired our contemplation of extending the domain of wavelet analysis into the realm of two-dimensional results, intended for the exploration of 2DFS in scientific inquiry. It is worth noting that similar initiatives have been undertaken by others, as exemplified by the functions within {\tt Python's pywavelet} \citep{2019JOSS....4.1237L} library. However, our distinctive aspiration revolves around harnessing the continuity inherent in wavelets to dissect pulse stacks exhibiting evolving drift patterns, with the objective of attaining lucid insights of variations.

\section*{Acknowledgements}
We express our gratitude to the reviewer for providing valuable suggestions that improve the manuscript. This work is supported by the the NSFC (No. 12133007) and National Key Research and Development Program of China (Grants No. 2021YFA0718503).

\appendix

\section{Appendix Fitting and $P_3$ Analysis}
\label{sec:appendix}
\begin{figure}[htbp]
    \centering
    \includegraphics[width=1.\linewidth]{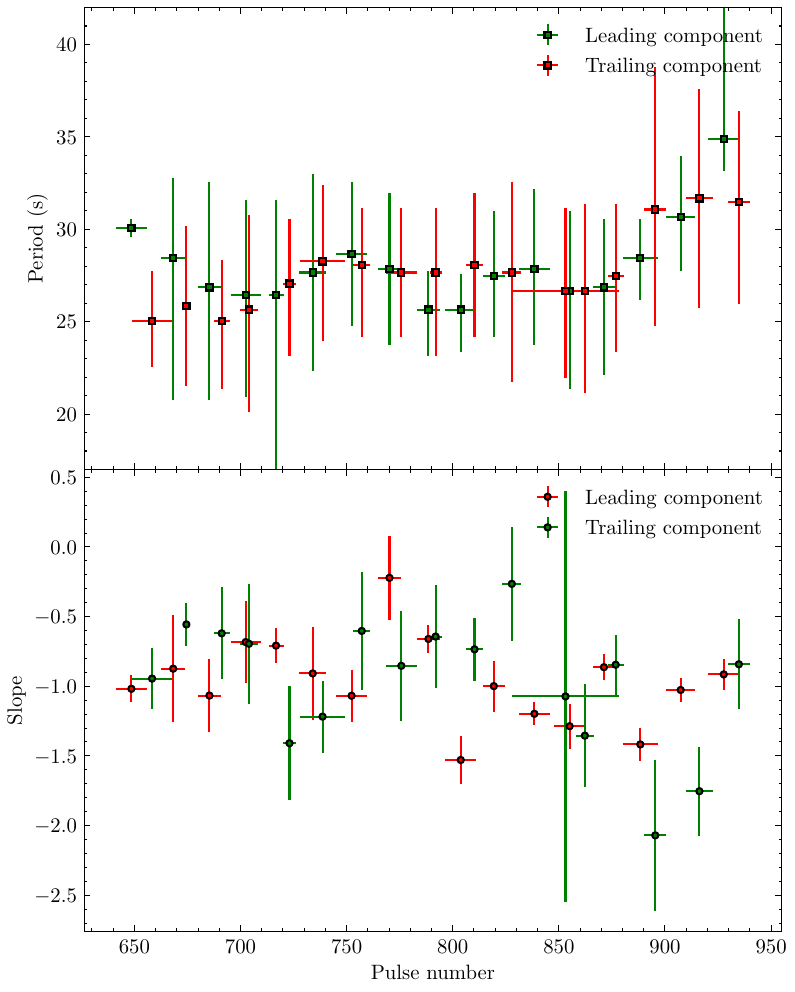}
    \caption{The temporal evolution of sub-pulse modulation periods and drifting rates. It is clearly that the period of sub-pulse modulation is anti-correlated with the drifting rate.}
    \label{fig:Periodic-slope}
\end{figure}

When compared with conventional FFT approaches, the waveley analysis method exhibits heightened advantages in assessing the temporal and frequency evolution of periods, thereby promoting a more extensive and detailed analytical framework. For example, we can extract the frequency information of isolated sub-pulses from the wavelet analysis results, As shown in Figure \ref{fig:Slope-F-fitting.pdf} and Figure \ref{fig:Periodic-slope}. Figure \ref{fig:Periodic-slope} illustrates a distinct anti-correlation between the sub-pulse modulation period and drifting rate, a relationship that aligns significantly with the robust fitting depicted in Figure \ref{fig:Slope-F-fitting.pdf}. 

\begin{figure}[htbp]
    \centering
    \includegraphics[width=1.\linewidth]{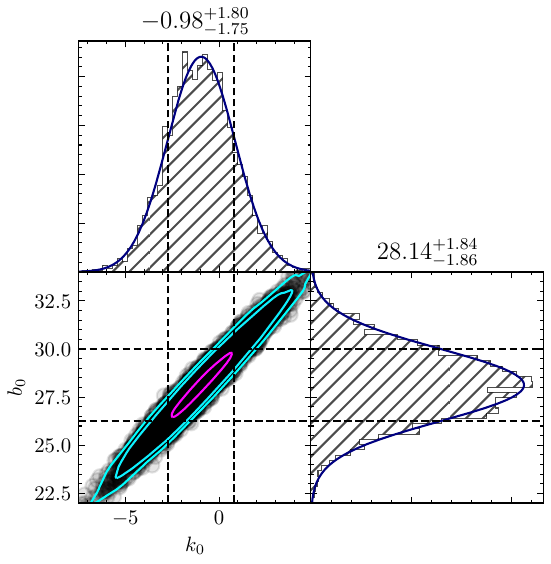}
    \caption{The corner drawing of the two coupling parameters of the fitting result in Figure \ref{fig:Slope-F-fitting.pdf}. $k_0$ and $b_0$ are the slope and intercept of the best linear fitting respectively. While the blue solid lines in the top and right panels indicate the norm fitting.}
    \label{fig:corner}
\end{figure}

The bottom panel of Figure \ref{fig:Periodic-slope} shows a more disarrayed and irregular result in comparison to that in the top panel. This discrepancy may primarily stem from the inadequacy of the employed method for calculation of the drifting rate and the insufficient clarity in the structural composition of some sub-pulses, as evident in points in the result of Figure \ref{fig:Periodic-slope} where some of drifting rates are almost zero. Fortunately, such results are sufficient for us to determine the negative correlation between the period and drifting rate.

In Figure \ref{fig:corner}, the coherence of the two parameters during the fitting process is substantiated, given that the drift rate exhibits a nearly inverse proportionality to the reciprocal of $P_3$ with a correlation coefficient approaching -1. Consequently, we confidently assert that there is a well-defined linear relationship of approximately -0.98 between the sub-pulse modulation period and drifting rate. The intercept $b_0$, however, aligns with the period when the drifting rate is zero, thereby substantiating the proposition that the intrinsic period of the sub-pulses is 28.14$^{+1.84}_{-1.86}$ s.

\begin{figure}[htbp]
    \centering
    \includegraphics[width=1.\linewidth]{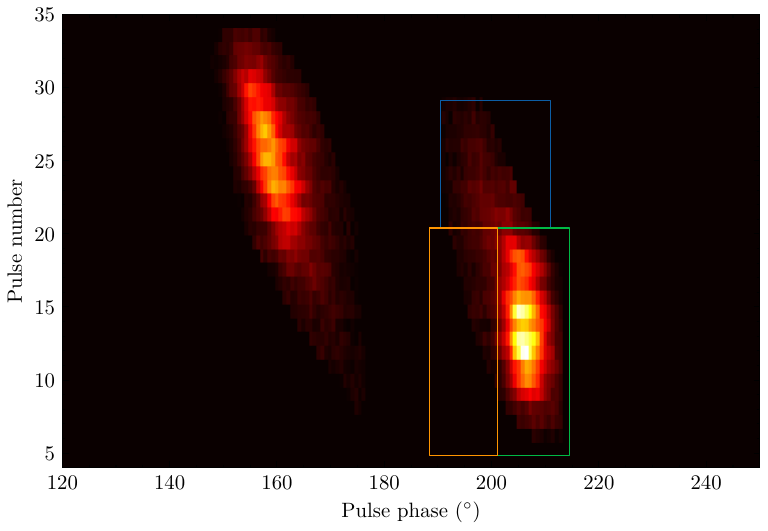}
    \caption{The result of P3 folding with only one sub-pulse preserved from both the leading and trailing components. The blue, orange and green squares show the two parts with different phase variations of tailing component..}
    \label{fig:p3folding_new}
\end{figure}

Another interesting thing is that within the results of $P_3$ folding (The algorithm used to the $P_3$ folding is the pulsar folding method \citep{2004hpa..book.....L}), as shown in Figure \ref{fig:p3folding} and \ref{fig:p3folding_new}, it appears that there are two different drifting behaviors present in the sub-pulses of the trailing component, which, in fact, are nearly identical. As depicted in Figure \ref{fig:p3folding_new}, when considering the independent sub-pulses from the two components within the results of $P_3$ folding, we noticed that the drifting behavior of the sub-pulses in the trailing component is essentially same as that of the sub-pulses in the leading component. However, certain portion of the sub-pulses in the trailing component exhibit relatively weaker intensities (for example, the certain part of the trailing component's sub-pulse in Figure 1, enclosed by the orange box, shares the same phase as C3), thereby resulting in different phase variations of the sub-pulses at different phases. Therefore, we postulate that the anomaly part of the sub-pulse in trailing component perhaps lies only within the green-boxed region (C4) of the trailing component. In actuality, regardless of whether it is the leading or trailing component, the drifting behavior of their sub-pulses should be consistent (linear), with variations observed only in certain sub-pulse components' drifting behaviors.

\begin{figure*}
\centering
\includegraphics[width=.48\textwidth]{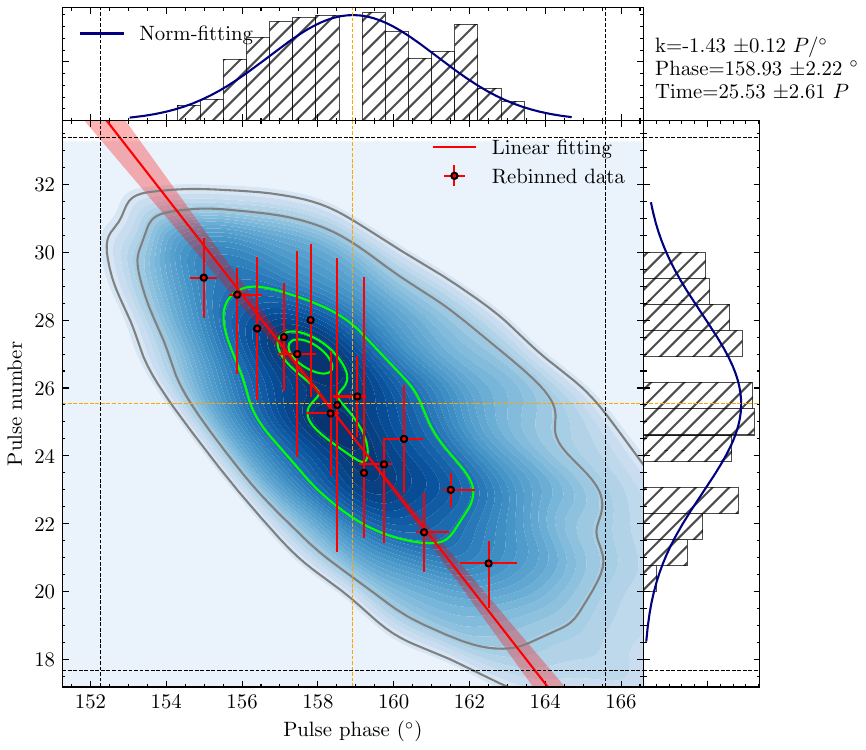}
\includegraphics[width=.48\textwidth]{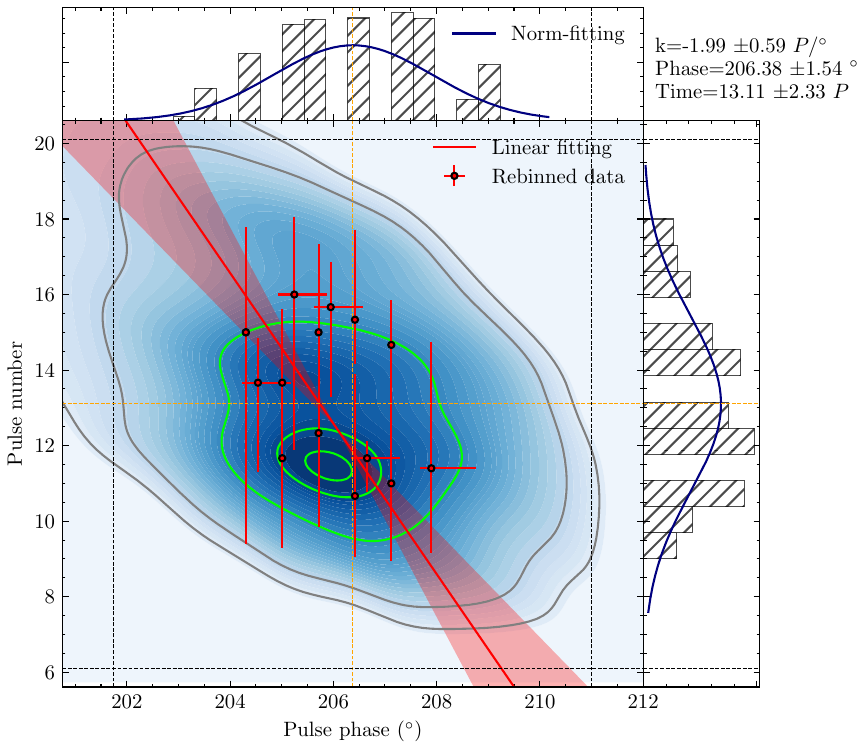}
\includegraphics[width=.48\textwidth]{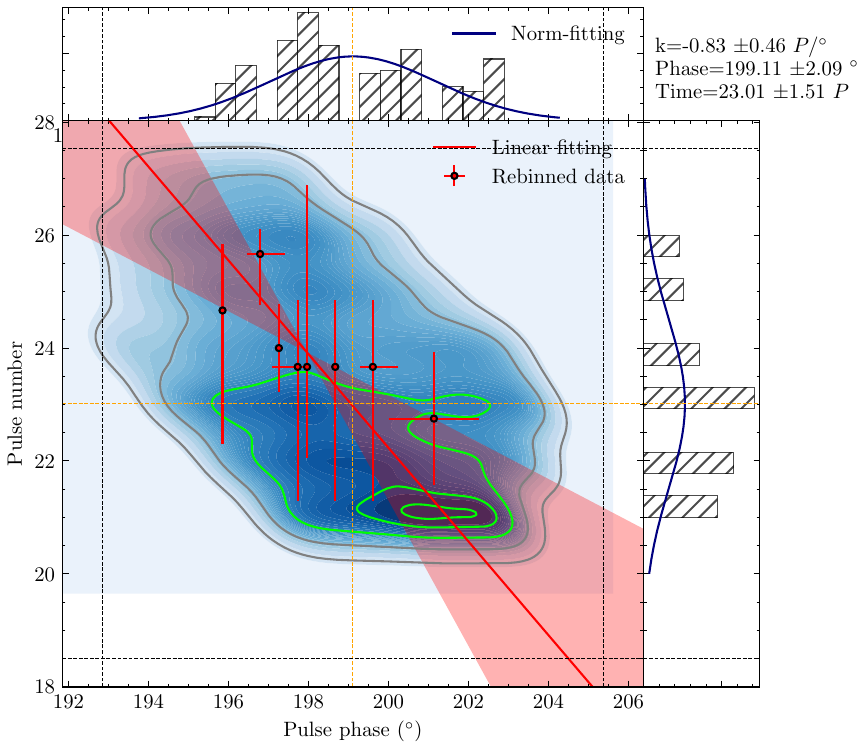}
\includegraphics[width=.48\textwidth]{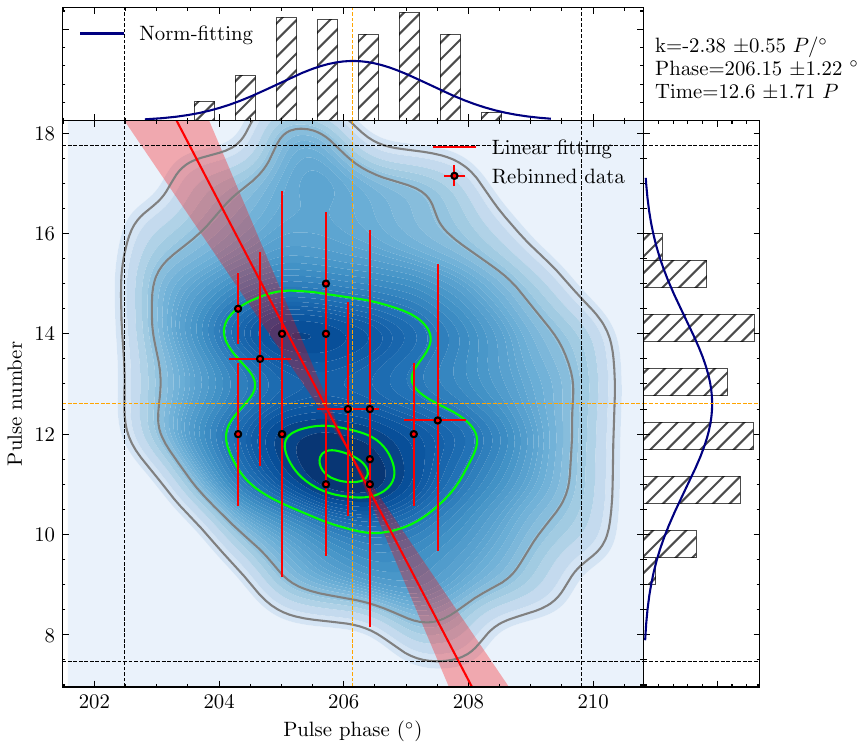}
\caption{Top: The contour plots of sub-pulses in the leading (left) and trailing components (right) in Figure \ref{fig:p3folding_new}. Bottom: the contour plots of regions enclosed by blue (left) and green square (right) in Figure \ref{fig:p3folding_new}, whose phase are corresponding that of C3 and C4 components. }
\label{fig:contourplots}
\end{figure*}

In Figure \ref{fig:contourplots}, we are displaying the contour plots of the sub-pulse and its different parts of trialing component in $P_3$ folding result. The results of these contour plots are suggesting that the drifting rate of the region with green square in Figure \ref{fig:p3folding_new} is larger than that of single sub-pulse in leading or trailing component. On the contrary, the drifting rate of the region enclosed by the blue square in Figure \ref{fig:p3folding_new} is smaller. These facts indicate that, concerning the sub-pulses of the trailing component in the results of $P_3$ folding, these two regions indeed exhibit different drifting behaviors and independently from each other. However, when considering the orange square region, we thought that the sub-pulse of trailing component is not composed by that two regions with different drifting rates, but maybe a integrate sub-pulse with asymmetrical intensity distribution. As a result, the drifting rates of individual sub-pulses in the leading and trailing components are generally consistent.



\bibliographystyle{harv}

\bibliography{cas-refs}

\begin{thebibliography}{77}
\expandafter\ifx\csname natexlab\endcsname\relax\def\natexlab#1{#1}\fi
\providecommand{\url}[1]{\texttt{#1}}
\providecommand{\href}[2]{#2}
\providecommand{\path}[1]{#1}
\providecommand{\DOIprefix}{doi:}
\providecommand{\ArXivprefix}{arXiv:}
\providecommand{\URLprefix}{URL: }
\providecommand{\Pubmedprefix}{pmid:}
\providecommand{\doi}[1]{\href{http://dx.doi.org/#1}{\path{#1}}}
\providecommand{\Pubmed}[1]{\href{pmid:#1}{\path{#1}}}
\providecommand{\bibinfo}[2]{#2}
\ifx\xfnm\relax \def\xfnm[#1]{\unskip,\space#1}\fi
\bibitem[{{Backer}(1970a)}]{1970Natur.228..752B}
\bibinfo{author}{{Backer}, D.C.}, \bibinfo{year}{1970}a.
\newblock \bibinfo{title}{{Correlated Subpulse Structure in PSR 1237 + 25}}.
\newblock \bibinfo{journal}{\nat} \bibinfo{volume}{228},
  \bibinfo{pages}{752--755}.
\newblock \DOIprefix\doi{10.1038/228752a0}.
\bibitem[{{Backer}(1970b)}]{1970Natur.227..692B}
\bibinfo{author}{{Backer}, D.C.}, \bibinfo{year}{1970}b.
\newblock \bibinfo{title}{{Correlation of Subpulse Structure in a Sequence of
  Pulses from Pulsar PSR 1919+21}}.
\newblock \bibinfo{journal}{\nat} \bibinfo{volume}{227},
  \bibinfo{pages}{692--695}.
\newblock \DOIprefix\doi{10.1038/227692a0}.
\bibitem[{{Backer}(1970c)}]{1970Natur.228...42B}
\bibinfo{author}{{Backer}, D.C.}, \bibinfo{year}{1970}c.
\newblock \bibinfo{title}{{Pulsar Nulling Phenomena}}.
\newblock \bibinfo{journal}{\nat} \bibinfo{volume}{228},
  \bibinfo{pages}{42--43}.
\newblock \DOIprefix\doi{10.1038/228042a0}.
\bibitem[{{Backer}(1973)}]{1973ApJ...182..245B}
\bibinfo{author}{{Backer}, D.C.}, \bibinfo{year}{1973}.
\newblock \bibinfo{title}{{Pulsar Fluctuation Spectra and the Generalized
  Drifting-Subpulse Phenomenon}}.
\newblock \bibinfo{journal}{\apj} \bibinfo{volume}{182},
  \bibinfo{pages}{245--276}.
\newblock \DOIprefix\doi{10.1086/152134}.
\bibitem[{{Basu} et~al.(2022){Basu}, {Melikidze} and
  {Mitra}}]{2022ApJ...936...35B}
\bibinfo{author}{{Basu}, R.}, \bibinfo{author}{{Melikidze}, G.I.},
  \bibinfo{author}{{Mitra}, D.}, \bibinfo{year}{2022}.
\newblock \bibinfo{title}{{Two-dimensional Configuration and Temporal Evolution
  of Spark Discharges in Pulsars}}.
\newblock \bibinfo{journal}{\apj} \bibinfo{volume}{936}, \bibinfo{pages}{35}.
\newblock \DOIprefix\doi{10.3847/1538-4357/ac8479},
  \href{http://arxiv.org/abs/2207.12908}{{\tt arXiv:2207.12908}}.
\bibitem[{{Basu} et~al.(2017){Basu}, {Mitra} and
  {Melikidze}}]{2017ApJ...846..109B}
\bibinfo{author}{{Basu}, R.}, \bibinfo{author}{{Mitra}, D.},
  \bibinfo{author}{{Melikidze}, G.I.}, \bibinfo{year}{2017}.
\newblock \bibinfo{title}{{Meterwavelength Single-pulse Polarimetric Emission
  Survey. III. The Phenomenon of Nulling in Pulsars}}.
\newblock \bibinfo{journal}{\apj} \bibinfo{volume}{846}, \bibinfo{pages}{109}.
\newblock \DOIprefix\doi{10.3847/1538-4357/aa862d},
  \href{http://arxiv.org/abs/1708.02499}{{\tt arXiv:1708.02499}}.
\bibitem[{{Basu} et~al.(2020a){Basu}, {Mitra} and
  {Melikidze}}]{2020MNRAS.496..465B}
\bibinfo{author}{{Basu}, R.}, \bibinfo{author}{{Mitra}, D.},
  \bibinfo{author}{{Melikidze}, G.I.}, \bibinfo{year}{2020}a.
\newblock \bibinfo{title}{{A mechanism of spark motion in inner acceleration
  region to investigate subpulse drifting in pulsars}}.
\newblock \bibinfo{journal}{\mnras} \bibinfo{volume}{496},
  \bibinfo{pages}{465--482}.
\newblock \DOIprefix\doi{10.1093/mnras/staa1574},
  \href{http://arxiv.org/abs/2006.01788}{{\tt arXiv:2006.01788}}.
\bibitem[{{Basu} et~al.(2020b){Basu}, {Mitra} and
  {Melikidze}}]{2020ApJ...889..133B}
\bibinfo{author}{{Basu}, R.}, \bibinfo{author}{{Mitra}, D.},
  \bibinfo{author}{{Melikidze}, G.I.}, \bibinfo{year}{2020}b.
\newblock \bibinfo{title}{{Periodic Modulation: Newly Emergent Emission
  Behavior in Pulsars}}.
\newblock \bibinfo{journal}{\apj} \bibinfo{volume}{889}, \bibinfo{pages}{133}.
\newblock \DOIprefix\doi{10.3847/1538-4357/ab63c9},
  \href{http://arxiv.org/abs/1912.06868}{{\tt arXiv:1912.06868}}.
\bibitem[{{Basu} et~al.(2023){Basu}, {Mitra} and
  {Melikidze}}]{2023ApJ...959...92B}
\bibinfo{author}{{Basu}, R.}, \bibinfo{author}{{Mitra}, D.},
  \bibinfo{author}{{Melikidze}, G.I.}, \bibinfo{year}{2023}.
\newblock \bibinfo{title}{{Mode Changing in PSR B0844-35 and PSR B1758-29 with
  Enhanced Emission at the Profile Centers}}.
\newblock \bibinfo{journal}{\apj} \bibinfo{volume}{959}, \bibinfo{pages}{92}.
\newblock \DOIprefix\doi{10.3847/1538-4357/ad083d},
  \href{http://arxiv.org/abs/2310.19725}{{\tt arXiv:2310.19725}}.
\bibitem[{{Basu} et~al.(2016){Basu}, {Mitra}, {Melikidze}, {Maciesiak},
  {Skrzypczak} and {Szary}}]{2016ApJ...833...29B}
\bibinfo{author}{{Basu}, R.}, \bibinfo{author}{{Mitra}, D.},
  \bibinfo{author}{{Melikidze}, G.I.}, \bibinfo{author}{{Maciesiak}, K.},
  \bibinfo{author}{{Skrzypczak}, A.}, \bibinfo{author}{{Szary}, A.},
  \bibinfo{year}{2016}.
\newblock \bibinfo{title}{{Meterwavelength Single-pulse Polarimetric Emission
  Survey. II. the Phenomenon of Drifting Subpulses}}.
\newblock \bibinfo{journal}{\apj} \bibinfo{volume}{833}, \bibinfo{pages}{29}.
\newblock \DOIprefix\doi{10.3847/1538-4357/833/1/29},
  \href{http://arxiv.org/abs/1608.00050}{{\tt arXiv:1608.00050}}.
\bibitem[{{Basu} et~al.(2019){Basu}, {Mitra}, {Melikidze} and
  {Skrzypczak}}]{2019MNRAS.482.3757B}
\bibinfo{author}{{Basu}, R.}, \bibinfo{author}{{Mitra}, D.},
  \bibinfo{author}{{Melikidze}, G.I.}, \bibinfo{author}{{Skrzypczak}, A.},
  \bibinfo{year}{2019}.
\newblock \bibinfo{title}{{Classification of subpulse drifting in pulsars}}.
\newblock \bibinfo{journal}{\mnras} \bibinfo{volume}{482},
  \bibinfo{pages}{3757--3788}.
\newblock \DOIprefix\doi{10.1093/mnras/sty2846},
  \href{http://arxiv.org/abs/1810.08423}{{\tt arXiv:1810.08423}}.
\bibitem[{{Bhat} et~al.(2007){Bhat}, {Gupta}, {Kramer}, {Karastergiou}, {Lyne}
  and {Johnston}}]{2007A&A...462..257B}
\bibinfo{author}{{Bhat}, N.D.R.}, \bibinfo{author}{{Gupta}, Y.},
  \bibinfo{author}{{Kramer}, M.}, \bibinfo{author}{{Karastergiou}, A.},
  \bibinfo{author}{{Lyne}, A.G.}, \bibinfo{author}{{Johnston}, S.},
  \bibinfo{year}{2007}.
\newblock \bibinfo{title}{{Simultaneous single-pulse observations of radio
  pulsars. V. On the broadband nature of the pulse nulling phenomenon in PSR
  B1133+16}}.
\newblock \bibinfo{journal}{\aap} \bibinfo{volume}{462},
  \bibinfo{pages}{257--268}.
\newblock \DOIprefix\doi{10.1051/0004-6361:20053157},
  \href{http://arxiv.org/abs/astro-ph/0610929}{{\tt arXiv:astro-ph/0610929}}.
\bibitem[{{Bhattacharyya} et~al.(2009){Bhattacharyya}, {Gupta} and
  {Gil}}]{2009MNRAS.398.1435B}
\bibinfo{author}{{Bhattacharyya}, B.}, \bibinfo{author}{{Gupta}, Y.},
  \bibinfo{author}{{Gil}, J.}, \bibinfo{year}{2009}.
\newblock \bibinfo{title}{{Exploring the remarkable subpulse drift and
  polarization properties of PSR B0818-41}}.
\newblock \bibinfo{journal}{\mnras} \bibinfo{volume}{398},
  \bibinfo{pages}{1435--1449}.
\newblock \DOIprefix\doi{10.1111/j.1365-2966.2009.15210.x},
  \href{http://arxiv.org/abs/0810.5227}{{\tt arXiv:0810.5227}}.
\bibitem[{{Biggs} et~al.(1985){Biggs}, {McCulloch}, {Hamilton}, {Manchester}
  and {Lyne}}]{1985MNRAS.215..281B}
\bibinfo{author}{{Biggs}, J.D.}, \bibinfo{author}{{McCulloch}, P.M.},
  \bibinfo{author}{{Hamilton}, P.A.}, \bibinfo{author}{{Manchester}, R.N.},
  \bibinfo{author}{{Lyne}, A.G.}, \bibinfo{year}{1985}.
\newblock \bibinfo{title}{{A study of PSR 0826-34 - a remarkable pulsar.}}
\newblock \bibinfo{journal}{\mnras} \bibinfo{volume}{215},
  \bibinfo{pages}{281--294}.
\newblock \DOIprefix\doi{10.1093/mnras/215.2.281}.
\bibitem[{{Chen} et~al.(2022a){Chen}, {Wang}, {Tian}, {Zhang}, {Liu}, {Wu},
  {Sai}, {Huang}, {Song}, {Qu}, {Tao}, {Zhang}, {Lu} and
  {Zhang}}]{2022MNRAS.517..182C}
\bibinfo{author}{{Chen}, X.}, \bibinfo{author}{{Wang}, W.},
  \bibinfo{author}{{Tian}, P.F.}, \bibinfo{author}{{Zhang}, P.},
  \bibinfo{author}{{Liu}, Q.}, \bibinfo{author}{{Wu}, H.J.},
  \bibinfo{author}{{Sai}, N.}, \bibinfo{author}{{Huang}, Y.},
  \bibinfo{author}{{Song}, L.M.}, \bibinfo{author}{{Qu}, J.L.},
  \bibinfo{author}{{Tao}, L.}, \bibinfo{author}{{Zhang}, S.},
  \bibinfo{author}{{Lu}, F.J.}, \bibinfo{author}{{Zhang}, S.N.},
  \bibinfo{year}{2022}a.
\newblock \bibinfo{title}{{Wavelet analysis of the transient QPOs in MAXI
  J1535-571 with Insight-HXMT}}.
\newblock \bibinfo{journal}{\mnras} \bibinfo{volume}{517},
  \bibinfo{pages}{182--191}.
\newblock \DOIprefix\doi{10.1093/mnras/stac2710},
  \href{http://arxiv.org/abs/2209.10408}{{\tt arXiv:2209.10408}}.
\bibitem[{{Chen} et~al.(2022b){Chen}, {Wang}, {You}, {Tian}, {Liu}, {Zhang},
  {Ding}, {Qu}, {Zhang}, {Song}, {Lu} and {Zhang}}]{2022MNRAS.513.4875C}
\bibinfo{author}{{Chen}, X.}, \bibinfo{author}{{Wang}, W.},
  \bibinfo{author}{{You}, B.}, \bibinfo{author}{{Tian}, P.F.},
  \bibinfo{author}{{Liu}, Q.}, \bibinfo{author}{{Zhang}, P.},
  \bibinfo{author}{{Ding}, Y.Z.}, \bibinfo{author}{{Qu}, J.L.},
  \bibinfo{author}{{Zhang}, S.N.}, \bibinfo{author}{{Song}, L.M.},
  \bibinfo{author}{{Lu}, F.J.}, \bibinfo{author}{{Zhang}, S.},
  \bibinfo{year}{2022}b.
\newblock \bibinfo{title}{{Wavelet analysis of MAXI J1535-571 with
  Insight-HXMT}}.
\newblock \bibinfo{journal}{\mnras} \bibinfo{volume}{513},
  \bibinfo{pages}{4875--4886}.
\newblock \DOIprefix\doi{10.1093/mnras/stac1175},
  \href{http://arxiv.org/abs/2204.12030}{{\tt arXiv:2204.12030}}.
\bibitem[{{Chen} et~al.(2023){Chen}, {Yan}, {Han}, {Wang}, {Wang}, {Jing},
  {Lee}, {Zhang}, {Xu}, {Wang}, {Yang}, {Su}, {Cai}, {Wang}, {Qiao}, {Xu} and
  {Zhou}}]{2023NatAs.tmp..177C}
\bibinfo{author}{{Chen}, X.}, \bibinfo{author}{{Yan}, Y.},
  \bibinfo{author}{{Han}, J.L.}, \bibinfo{author}{{Wang}, C.},
  \bibinfo{author}{{Wang}, P.F.}, \bibinfo{author}{{Jing}, W.C.},
  \bibinfo{author}{{Lee}, K.J.}, \bibinfo{author}{{Zhang}, B.},
  \bibinfo{author}{{Xu}, R.X.}, \bibinfo{author}{{Wang}, T.},
  \bibinfo{author}{{Yang}, Z.L.}, \bibinfo{author}{{Su}, W.Q.},
  \bibinfo{author}{{Cai}, N.N.}, \bibinfo{author}{{Wang}, W.Y.},
  \bibinfo{author}{{Qiao}, G.J.}, \bibinfo{author}{{Xu}, J.},
  \bibinfo{author}{{Zhou}, D.J.}, \bibinfo{year}{2023}.
\newblock \bibinfo{title}{{Strong and weak pulsar radio emission due to
  thunderstorms and raindrops of particles in the magnetosphere}}.
\newblock \bibinfo{journal}{Nature Astronomy}
  \DOIprefix\doi{10.1038/s41550-023-02056-z},
  \href{http://arxiv.org/abs/2306.12017}{{\tt arXiv:2306.12017}}.
\bibitem[{{Cheng} and {Ruderman}(1980)}]{1980ApJ...235..576C}
\bibinfo{author}{{Cheng}, A.F.}, \bibinfo{author}{{Ruderman}, M.A.},
  \bibinfo{year}{1980}.
\newblock \bibinfo{title}{{Particle acceleration and radio emission above
  pulsar polar caps.}}
\newblock \bibinfo{journal}{\apj} \bibinfo{volume}{235},
  \bibinfo{pages}{576--586}.
\newblock \DOIprefix\doi{10.1086/157661}.
\bibitem[{{Coles} et~al.(2010){Coles}, {Rickett}, {Gao}, {Hobbs} and
  {Verbiest}}]{2010ApJ...717.1206C}
\bibinfo{author}{{Coles}, W.A.}, \bibinfo{author}{{Rickett}, B.J.},
  \bibinfo{author}{{Gao}, J.J.}, \bibinfo{author}{{Hobbs}, G.},
  \bibinfo{author}{{Verbiest}, J.P.W.}, \bibinfo{year}{2010}.
\newblock \bibinfo{title}{{Scattering of Pulsar Radio Emission by the
  Interstellar Plasma}}.
\newblock \bibinfo{journal}{\apj} \bibinfo{volume}{717},
  \bibinfo{pages}{1206--1221}.
\newblock \DOIprefix\doi{10.1088/0004-637X/717/2/1206},
  \href{http://arxiv.org/abs/1005.4914}{{\tt arXiv:1005.4914}}.
\bibitem[{{De Moortel} et~al.(2004){De Moortel}, {Munday} and
  {Hood}}]{2004SoPh..222..203D}
\bibinfo{author}{{De Moortel}, I.}, \bibinfo{author}{{Munday}, S.A.},
  \bibinfo{author}{{Hood}, A.W.}, \bibinfo{year}{2004}.
\newblock \bibinfo{title}{{Wavelet Analysis: the effect of varying basic
  wavelet parameters}}.
\newblock \bibinfo{journal}{\solphys} \bibinfo{volume}{222},
  \bibinfo{pages}{203--228}.
\newblock \DOIprefix\doi{10.1023/B:SOLA.0000043578.01201.2d}.
\bibitem[{{Deshpande} and {Rankin}(1999)}]{1999ApJ...524.1008D}
\bibinfo{author}{{Deshpande}, A.A.}, \bibinfo{author}{{Rankin}, J.M.},
  \bibinfo{year}{1999}.
\newblock \bibinfo{title}{{Pulsar Magnetospheric Emission Mapping: Images and
  Implications of Polar CAP Weather}}.
\newblock \bibinfo{journal}{\apj} \bibinfo{volume}{524},
  \bibinfo{pages}{1008--1013}.
\newblock \DOIprefix\doi{10.1086/307862},
  \href{http://arxiv.org/abs/astro-ph/9909398}{{\tt arXiv:astro-ph/9909398}}.
\bibitem[{{Deshpande} and {Rankin}(2001)}]{2001MNRAS.322..438D}
\bibinfo{author}{{Deshpande}, A.A.}, \bibinfo{author}{{Rankin}, J.M.},
  \bibinfo{year}{2001}.
\newblock \bibinfo{title}{{The topology and polarization of sub-beams
  associated with the `drifting' sub-pulse emission of pulsar B0943+10 - I.
  Analysis of Arecibo 430- and 111-MHz observations}}.
\newblock \bibinfo{journal}{\mnras} \bibinfo{volume}{322},
  \bibinfo{pages}{438--460}.
\newblock \DOIprefix\doi{10.1046/j.1365-8711.2001.04079.x},
  \href{http://arxiv.org/abs/astro-ph/0010048}{{\tt arXiv:astro-ph/0010048}}.
\bibitem[{{Drake} and {Craft}(1968)}]{1968Natur.220..231D}
\bibinfo{author}{{Drake}, F.D.}, \bibinfo{author}{{Craft}, H.D.},
  \bibinfo{year}{1968}.
\newblock \bibinfo{title}{{Second Periodic Pulsation in Pulsars}}.
\newblock \bibinfo{journal}{\nat} \bibinfo{volume}{220},
  \bibinfo{pages}{231--235}.
\newblock \DOIprefix\doi{10.1038/220231a0}.
\bibitem[{{Edwards} and {Stappers}(2002)}]{2002A&A...393..733E}
\bibinfo{author}{{Edwards}, R.T.}, \bibinfo{author}{{Stappers}, B.W.},
  \bibinfo{year}{2002}.
\newblock \bibinfo{title}{{Drifting sub-pulse analysis using the
  two-dimensional Fourier transform}}.
\newblock \bibinfo{journal}{\aap} \bibinfo{volume}{393},
  \bibinfo{pages}{733--748}.
\newblock \DOIprefix\doi{10.1051/0004-6361:20021067},
  \href{http://arxiv.org/abs/astro-ph/0207472}{{\tt arXiv:astro-ph/0207472}}.
\bibitem[{{Edwards} and {Stappers}(2004)}]{2004A&A...421..681E}
\bibinfo{author}{{Edwards}, R.T.}, \bibinfo{author}{{Stappers}, B.W.},
  \bibinfo{year}{2004}.
\newblock \bibinfo{title}{{Ellipticity and deviations from orthogonality in the
  polarization modes of PSR B0329+54}}.
\newblock \bibinfo{journal}{\aap} \bibinfo{volume}{421},
  \bibinfo{pages}{681--691}.
\newblock \DOIprefix\doi{10.1051/0004-6361:20040228},
  \href{http://arxiv.org/abs/astro-ph/0404092}{{\tt arXiv:astro-ph/0404092}}.
\bibitem[{{Esamdin} et~al.(2005){Esamdin}, {Lyne}, {Graham-Smith}, {Kramer},
  {Manchester} and {Wu}}]{2005MNRAS.356...59E}
\bibinfo{author}{{Esamdin}, A.}, \bibinfo{author}{{Lyne}, A.G.},
  \bibinfo{author}{{Graham-Smith}, F.}, \bibinfo{author}{{Kramer}, M.},
  \bibinfo{author}{{Manchester}, R.N.}, \bibinfo{author}{{Wu}, X.},
  \bibinfo{year}{2005}.
\newblock \bibinfo{title}{{Mode switching and subpulse drifting in PSR
  B0826-34}}.
\newblock \bibinfo{journal}{\mnras} \bibinfo{volume}{356},
  \bibinfo{pages}{59--65}.
\newblock \DOIprefix\doi{10.1111/j.1365-2966.2004.08444.x},
  \href{http://arxiv.org/abs/astro-ph/0410018}{{\tt arXiv:astro-ph/0410018}}.
\bibitem[{{Farge}(1992)}]{1992AnRFM..24..395F}
\bibinfo{author}{{Farge}, M.}, \bibinfo{year}{1992}.
\newblock \bibinfo{title}{{Wavelet transforms and their applications to
  turbulence}}.
\newblock \bibinfo{journal}{Annual Review of Fluid Mechanics}
  \bibinfo{volume}{24}, \bibinfo{pages}{395--457}.
\newblock \DOIprefix\doi{10.1146/annurev.fl.24.010192.002143}.
\bibitem[{{Filippenko} and {Radhakrishnan}(1982)}]{1982ApJ...263..828F}
\bibinfo{author}{{Filippenko}, A.V.}, \bibinfo{author}{{Radhakrishnan}, V.},
  \bibinfo{year}{1982}.
\newblock \bibinfo{title}{{Pulsar nulling and drifting subpulse memory.}}
\newblock \bibinfo{journal}{\apj} \bibinfo{volume}{263},
  \bibinfo{pages}{828--834}.
\newblock \DOIprefix\doi{10.1086/160553}.
\bibitem[{{Foreman-Mackey} et~al.(2013){Foreman-Mackey}, {Hogg}, {Lang} and
  {Goodman}}]{2013PASP..125..306F}
\bibinfo{author}{{Foreman-Mackey}, D.}, \bibinfo{author}{{Hogg}, D.W.},
  \bibinfo{author}{{Lang}, D.}, \bibinfo{author}{{Goodman}, J.},
  \bibinfo{year}{2013}.
\newblock \bibinfo{title}{{emcee: The MCMC Hammer}}.
\newblock \bibinfo{journal}{\pasp} \bibinfo{volume}{125}, \bibinfo{pages}{306}.
\newblock \DOIprefix\doi{10.1086/670067},
  \href{http://arxiv.org/abs/1202.3665}{{\tt arXiv:1202.3665}}.
\bibitem[{{Gajjar} et~al.(2014){Gajjar}, {Joshi}, {Kramer}, {Karuppusamy} and
  {Smits}}]{2014ApJ...797...18G}
\bibinfo{author}{{Gajjar}, V.}, \bibinfo{author}{{Joshi}, B.C.},
  \bibinfo{author}{{Kramer}, M.}, \bibinfo{author}{{Karuppusamy}, R.},
  \bibinfo{author}{{Smits}, R.}, \bibinfo{year}{2014}.
\newblock \bibinfo{title}{{Frequency Independent Quenching of Pulsed
  Emission}}.
\newblock \bibinfo{journal}{\apj} \bibinfo{volume}{797}, \bibinfo{pages}{18}.
\newblock \DOIprefix\doi{10.1088/0004-637X/797/1/18},
  \href{http://arxiv.org/abs/1409.5589}{{\tt arXiv:1409.5589}}.
\bibitem[{{Gil} et~al.(2003){Gil}, {Melikidze} and
  {Geppert}}]{2003A&A...407..315G}
\bibinfo{author}{{Gil}, J.}, \bibinfo{author}{{Melikidze}, G.I.},
  \bibinfo{author}{{Geppert}, U.}, \bibinfo{year}{2003}.
\newblock \bibinfo{title}{{Drifting subpulses and inner accelerationregions in
  radio pulsars}}.
\newblock \bibinfo{journal}{\aap} \bibinfo{volume}{407},
  \bibinfo{pages}{315--324}.
\newblock \DOIprefix\doi{10.1051/0004-6361:20030854},
  \href{http://arxiv.org/abs/astro-ph/0305463}{{\tt arXiv:astro-ph/0305463}}.
\bibitem[{{Gil} and {Sendyk}(2000)}]{2000ApJ...541..351G}
\bibinfo{author}{{Gil}, J.A.}, \bibinfo{author}{{Sendyk}, M.},
  \bibinfo{year}{2000}.
\newblock \bibinfo{title}{{Spark Model for Pulsar Radiation Modulation
  Patterns}}.
\newblock \bibinfo{journal}{\apj} \bibinfo{volume}{541},
  \bibinfo{pages}{351--366}.
\newblock \DOIprefix\doi{10.1086/309394},
  \href{http://arxiv.org/abs/astro-ph/0002450}{{\tt arXiv:astro-ph/0002450}}.
\bibitem[{{Hassall} et~al.(2013){Hassall}, {Stappers}, {Weltevrede}, {Hessels},
  {Alexov}, {Coenen}, {Karastergiou}, {Kramer}, {Keane}, {Kondratiev}, {van
  Leeuwen}, {Noutsos}, {Pilia}, {Serylak}, {Sobey}, {Zagkouris}, {Fender},
  {Bell}, {Broderick}, {Eisl{\"o}ffel}, {Falcke}, {Grie{\ss}meier},
  {Kuniyoshi}, {Miller-Jones}, {Wise}, {Wucknitz}, {Zarka}, {Asgekar},
  {Batejat}, {Bentum}, {Bernardi}, {Best}, {Bonafede}, {Breitling},
  {Br{\"u}ggen}, {Butcher}, {Ciardi}, {de Gasperin}, {de Reijer}, {Duscha},
  {Fallows}, {Ferrari}, {Frieswijk}, {Garrett}, {Gunst}, {Heald}, {Hoeft},
  {Juette}, {Maat}, {McKean}, {Norden}, {Pandey-Pommier}, {Pizzo}, {Polatidis},
  {Reich}, {R{\"o}ttgering}, {Sluman}, {Tang}, {Tasse}, {Vermeulen}, {van
  Weeren}, {Wijnholds} and {Yatawatta}}]{2013A&A...552A..61H}
\bibinfo{author}{{Hassall}, T.E.}, \bibinfo{author}{{Stappers}, B.W.},
  \bibinfo{author}{{Weltevrede}, P.}, \bibinfo{author}{{Hessels}, J.W.T.},
  \bibinfo{author}{{Alexov}, A.}, \bibinfo{author}{{Coenen}, T.},
  \bibinfo{author}{{Karastergiou}, A.}, \bibinfo{author}{{Kramer}, M.},
  \bibinfo{author}{{Keane}, E.F.}, \bibinfo{author}{{Kondratiev}, V.I.},
  \bibinfo{author}{{van Leeuwen}, J.}, \bibinfo{author}{{Noutsos}, A.},
  \bibinfo{author}{{Pilia}, M.}, \bibinfo{author}{{Serylak}, M.},
  \bibinfo{author}{{Sobey}, C.}, \bibinfo{author}{{Zagkouris}, K.},
  \bibinfo{author}{{Fender}, R.}, \bibinfo{author}{{Bell}, M.E.},
  \bibinfo{author}{{Broderick}, J.}, \bibinfo{author}{{Eisl{\"o}ffel}, J.},
  \bibinfo{author}{{Falcke}, H.}, \bibinfo{author}{{Grie{\ss}meier}, J.M.},
  \bibinfo{author}{{Kuniyoshi}, M.}, \bibinfo{author}{{Miller-Jones}, J.C.A.},
  \bibinfo{author}{{Wise}, M.W.}, \bibinfo{author}{{Wucknitz}, O.},
  \bibinfo{author}{{Zarka}, P.}, \bibinfo{author}{{Asgekar}, A.},
  \bibinfo{author}{{Batejat}, F.}, \bibinfo{author}{{Bentum}, M.J.},
  \bibinfo{author}{{Bernardi}, G.}, \bibinfo{author}{{Best}, P.},
  \bibinfo{author}{{Bonafede}, A.}, \bibinfo{author}{{Breitling}, F.},
  \bibinfo{author}{{Br{\"u}ggen}, M.}, \bibinfo{author}{{Butcher}, H.R.},
  \bibinfo{author}{{Ciardi}, B.}, \bibinfo{author}{{de Gasperin}, F.},
  \bibinfo{author}{{de Reijer}, J.P.}, \bibinfo{author}{{Duscha}, S.},
  \bibinfo{author}{{Fallows}, R.A.}, \bibinfo{author}{{Ferrari}, C.},
  \bibinfo{author}{{Frieswijk}, W.}, \bibinfo{author}{{Garrett}, M.A.},
  \bibinfo{author}{{Gunst}, A.W.}, \bibinfo{author}{{Heald}, G.},
  \bibinfo{author}{{Hoeft}, M.}, \bibinfo{author}{{Juette}, E.},
  \bibinfo{author}{{Maat}, P.}, \bibinfo{author}{{McKean}, J.P.},
  \bibinfo{author}{{Norden}, M.J.}, \bibinfo{author}{{Pandey-Pommier}, M.},
  \bibinfo{author}{{Pizzo}, R.}, \bibinfo{author}{{Polatidis}, A.G.},
  \bibinfo{author}{{Reich}, W.}, \bibinfo{author}{{R{\"o}ttgering}, H.},
  \bibinfo{author}{{Sluman}, J.}, \bibinfo{author}{{Tang}, Y.},
  \bibinfo{author}{{Tasse}, C.}, \bibinfo{author}{{Vermeulen}, R.},
  \bibinfo{author}{{van Weeren}, R.J.}, \bibinfo{author}{{Wijnholds}, S.J.},
  \bibinfo{author}{{Yatawatta}, S.}, \bibinfo{year}{2013}.
\newblock \bibinfo{title}{{Differential frequency-dependent delay from the
  pulsar magnetosphere}}.
\newblock \bibinfo{journal}{\aap} \bibinfo{volume}{552}, \bibinfo{pages}{A61}.
\newblock \DOIprefix\doi{10.1051/0004-6361/201220764},
  \href{http://arxiv.org/abs/1302.2321}{{\tt arXiv:1302.2321}}.
\bibitem[{{Helfand} et~al.(1975){Helfand}, {Manchester} and
  {Taylor}}]{1975ApJ...198..661H}
\bibinfo{author}{{Helfand}, D.J.}, \bibinfo{author}{{Manchester}, R.N.},
  \bibinfo{author}{{Taylor}, J.H.}, \bibinfo{year}{1975}.
\newblock \bibinfo{title}{{Observations of pulsar radio emission. III.
  Stability of integrated profiles.}}
\newblock \bibinfo{journal}{\apj} \bibinfo{volume}{198},
  \bibinfo{pages}{661--670}.
\newblock \DOIprefix\doi{10.1086/153644}.
\bibitem[{Herrmann et~al.(2014)Herrmann, Rach, Vosskuhl and
  Str{\"u}ber}]{herrmann2014time}
\bibinfo{author}{Herrmann, C.S.}, \bibinfo{author}{Rach, S.},
  \bibinfo{author}{Vosskuhl, J.}, \bibinfo{author}{Str{\"u}ber, D.},
  \bibinfo{year}{2014}.
\newblock \bibinfo{title}{Time--frequency analysis of event-related potentials:
  a brief tutorial}.
\newblock \bibinfo{journal}{Brain topography} \bibinfo{volume}{27},
  \bibinfo{pages}{438--450}.
\bibitem[{{Hotan} et~al.(2004){Hotan}, {van Straten} and
  {Manchester}}]{2004PASA...21..302H}
\bibinfo{author}{{Hotan}, A.W.}, \bibinfo{author}{{van Straten}, W.},
  \bibinfo{author}{{Manchester}, R.N.}, \bibinfo{year}{2004}.
\newblock \bibinfo{title}{{PSRCHIVE and PSRFITS: An Open Approach to Radio
  Pulsar Data Storage and Analysis}}.
\newblock \bibinfo{journal}{\pasa} \bibinfo{volume}{21},
  \bibinfo{pages}{302--309}.
\newblock \DOIprefix\doi{10.1071/AS04022},
  \href{http://arxiv.org/abs/astro-ph/0404549}{{\tt arXiv:astro-ph/0404549}}.
\bibitem[{{Johnston} and {Weisberg}(2006)}]{2006MNRAS.368.1856J}
\bibinfo{author}{{Johnston}, S.}, \bibinfo{author}{{Weisberg}, J.M.},
  \bibinfo{year}{2006}.
\newblock \bibinfo{title}{{Profile morphology and polarization of young
  pulsars}}.
\newblock \bibinfo{journal}{\mnras} \bibinfo{volume}{368},
  \bibinfo{pages}{1856--1870}.
\newblock \DOIprefix\doi{10.1111/j.1365-2966.2006.10263.x},
  \href{http://arxiv.org/abs/astro-ph/0603037}{{\tt arXiv:astro-ph/0603037}}.
\bibitem[{{Kaiser} and {Hudgins}(1995)}]{1995PhT....48g..57K}
\bibinfo{author}{{Kaiser}, G.}, \bibinfo{author}{{Hudgins}, L.H.},
  \bibinfo{year}{1995}.
\newblock \bibinfo{title}{{A Friendly Guide to Wavelets}}.
\newblock \bibinfo{journal}{Physics Today} \bibinfo{volume}{48},
  \bibinfo{pages}{57}.
\newblock \DOIprefix\doi{10.1063/1.2808105}.
\bibitem[{{Karastergiou} and {Johnston}(2007)}]{2007MNRAS.380.1678K}
\bibinfo{author}{{Karastergiou}, A.}, \bibinfo{author}{{Johnston}, S.},
  \bibinfo{year}{2007}.
\newblock \bibinfo{title}{{An empirical model for the beams of radio pulsars}}.
\newblock \bibinfo{journal}{\mnras} \bibinfo{volume}{380},
  \bibinfo{pages}{1678--1684}.
\newblock \DOIprefix\doi{10.1111/j.1365-2966.2007.12237.x},
  \href{http://arxiv.org/abs/0707.2547}{{\tt arXiv:0707.2547}}.
\bibitem[{{Kaspi} and {Beloborodov}(2017)}]{2017ARA&A..55..261K}
\bibinfo{author}{{Kaspi}, V.M.}, \bibinfo{author}{{Beloborodov}, A.M.},
  \bibinfo{year}{2017}.
\newblock \bibinfo{title}{{Magnetars}}.
\newblock \bibinfo{journal}{\araa} \bibinfo{volume}{55},
  \bibinfo{pages}{261--301}.
\newblock \DOIprefix\doi{10.1146/annurev-astro-081915-023329},
  \href{http://arxiv.org/abs/1703.00068}{{\tt arXiv:1703.00068}}.
\bibitem[{{Keith} et~al.(2013){Keith}, {Shannon} and
  {Johnston}}]{2013MNRAS.432.3080K}
\bibinfo{author}{{Keith}, M.J.}, \bibinfo{author}{{Shannon}, R.M.},
  \bibinfo{author}{{Johnston}, S.}, \bibinfo{year}{2013}.
\newblock \bibinfo{title}{{A connection between radio state changing and glitch
  activity in PSR J0742-2822}}.
\newblock \bibinfo{journal}{\mnras} \bibinfo{volume}{432},
  \bibinfo{pages}{3080--3084}.
\newblock \DOIprefix\doi{10.1093/mnras/stt660},
  \href{http://arxiv.org/abs/1304.4644}{{\tt arXiv:1304.4644}}.
\bibitem[{{Kramer} et~al.(2006){Kramer}, {Lyne}, {O'Brien}, {Jordan} and
  {Lorimer}}]{2006Sci...312..549K}
\bibinfo{author}{{Kramer}, M.}, \bibinfo{author}{{Lyne}, A.G.},
  \bibinfo{author}{{O'Brien}, J.T.}, \bibinfo{author}{{Jordan}, C.A.},
  \bibinfo{author}{{Lorimer}, D.R.}, \bibinfo{year}{2006}.
\newblock \bibinfo{title}{{A Periodically Active Pulsar Giving Insight into
  Magnetospheric Physics}}.
\newblock \bibinfo{journal}{Science} \bibinfo{volume}{312},
  \bibinfo{pages}{549--551}.
\newblock \DOIprefix\doi{10.1126/science.1124060},
  \href{http://arxiv.org/abs/astro-ph/0604605}{{\tt arXiv:astro-ph/0604605}}.
\bibitem[{{Lee} et~al.(2019){Lee}, {Gommers}, {Waselewski}, {Wohlfahrt} and
  {O'Leary}}]{2019JOSS....4.1237L}
\bibinfo{author}{{Lee}, G.}, \bibinfo{author}{{Gommers}, R.},
  \bibinfo{author}{{Waselewski}, F.}, \bibinfo{author}{{Wohlfahrt}, K.},
  \bibinfo{author}{{O'Leary}, A.}, \bibinfo{year}{2019}.
\newblock \bibinfo{title}{{PyWavelets: A Python package for wavelet analysis}}.
\newblock \bibinfo{journal}{The Journal of Open Source Software}
  \bibinfo{volume}{4}, \bibinfo{pages}{1237}.
\newblock \DOIprefix\doi{10.21105/joss.01237}.
\bibitem[{{Lorimer} et~al.(2007){Lorimer}, {Bailes}, {McLaughlin}, {Narkevic}
  and {Crawford}}]{2007Sci...318..777L}
\bibinfo{author}{{Lorimer}, D.R.}, \bibinfo{author}{{Bailes}, M.},
  \bibinfo{author}{{McLaughlin}, M.A.}, \bibinfo{author}{{Narkevic}, D.J.},
  \bibinfo{author}{{Crawford}, F.}, \bibinfo{year}{2007}.
\newblock \bibinfo{title}{{A Bright Millisecond Radio Burst of Extragalactic
  Origin}}.
\newblock \bibinfo{journal}{Science} \bibinfo{volume}{318},
  \bibinfo{pages}{777}.
\newblock \DOIprefix\doi{10.1126/science.1147532},
  \href{http://arxiv.org/abs/0709.4301}{{\tt arXiv:0709.4301}}.
\bibitem[{{Lorimer} and {Kramer}(2004)}]{2004hpa..book.....L}
\bibinfo{author}{{Lorimer}, D.R.}, \bibinfo{author}{{Kramer}, M.},
  \bibinfo{year}{2004}.
\newblock \bibinfo{title}{{Handbook of Pulsar Astronomy}}.
  volume~\bibinfo{volume}{4}.
\bibitem[{{McSweeney} et~al.(2017){McSweeney}, {Bhat}, {Tremblay}, {Deshpande}
  and {Ord}}]{2017ApJ...836..224M}
\bibinfo{author}{{McSweeney}, S.J.}, \bibinfo{author}{{Bhat}, N.D.R.},
  \bibinfo{author}{{Tremblay}, S.E.}, \bibinfo{author}{{Deshpande}, A.A.},
  \bibinfo{author}{{Ord}, S.M.}, \bibinfo{year}{2017}.
\newblock \bibinfo{title}{{Low-frequency Observations of the Subpulse Drifter
  PSR J0034-0721 with the Murchison Widefield Array}}.
\newblock \bibinfo{journal}{\apj} \bibinfo{volume}{836}, \bibinfo{pages}{224}.
\newblock \DOIprefix\doi{10.3847/1538-4357/aa5c35},
  \href{http://arxiv.org/abs/1701.06755}{{\tt arXiv:1701.06755}}.
\bibitem[{{Mitra} and {Rankin}(2008)}]{2008MNRAS.385..606M}
\bibinfo{author}{{Mitra}, D.}, \bibinfo{author}{{Rankin}, J.M.},
  \bibinfo{year}{2008}.
\newblock \bibinfo{title}{{On the subpulse modulation, polarization and
  sub-beam carousel configuration of pulsar B1857-26}}.
\newblock \bibinfo{journal}{\mnras} \bibinfo{volume}{385},
  \bibinfo{pages}{606--613}.
\newblock \DOIprefix\doi{10.1111/j.1365-2966.2008.12838.x}.
\bibitem[{{Nyquist}(1928)}]{1928TAIEE..47..617N}
\bibinfo{author}{{Nyquist}, H.}, \bibinfo{year}{1928}.
\newblock \bibinfo{title}{{Certain Topics in Telegraph Transmission Theory}}.
\newblock \bibinfo{journal}{Transactions of the American Institute of
  Electrical Engineers} \bibinfo{volume}{47}, \bibinfo{pages}{617--624}.
\newblock \DOIprefix\doi{10.1109/T-AIEE.1928.5055024}.
\bibitem[{{Page}(1973)}]{1973MNRAS.163...29P}
\bibinfo{author}{{Page}, C.G.}, \bibinfo{year}{1973}.
\newblock \bibinfo{title}{{The drifting sub-pulse phenomenon in PSR 0809+74}}.
\newblock \bibinfo{journal}{\mnras} \bibinfo{volume}{163}, \bibinfo{pages}{29}.
\newblock \DOIprefix\doi{10.1093/mnras/163.1.29}.
\bibitem[{{Palfreyman} et~al.(2018){Palfreyman}, {Dickey}, {Hotan}, {Ellingsen}
  and {van Straten}}]{2018Natur.556..219P}
\bibinfo{author}{{Palfreyman}, J.}, \bibinfo{author}{{Dickey}, J.M.},
  \bibinfo{author}{{Hotan}, A.}, \bibinfo{author}{{Ellingsen}, S.},
  \bibinfo{author}{{van Straten}, W.}, \bibinfo{year}{2018}.
\newblock \bibinfo{title}{{Alteration of the magnetosphere of the Vela pulsar
  during a glitch}}.
\newblock \bibinfo{journal}{\nat} \bibinfo{volume}{556},
  \bibinfo{pages}{219--222}.
\newblock \DOIprefix\doi{10.1038/s41586-018-0001-x}.
\bibitem[{{Philippov} and {Kramer}(2022)}]{2022ARA&A..60..495P}
\bibinfo{author}{{Philippov}, A.}, \bibinfo{author}{{Kramer}, M.},
  \bibinfo{year}{2022}.
\newblock \bibinfo{title}{{Pulsar Magnetospheres and Their Radiation}}.
\newblock \bibinfo{journal}{\araa} \bibinfo{volume}{60},
  \bibinfo{pages}{495--558}.
\newblock \DOIprefix\doi{10.1146/annurev-astro-052920-112338}.
\bibitem[{{Poutanen}(2020)}]{2020A&A...641A.166P}
\bibinfo{author}{{Poutanen}, J.}, \bibinfo{year}{2020}.
\newblock \bibinfo{title}{{Relativistic rotating vector model for X-ray
  millisecond pulsars}}.
\newblock \bibinfo{journal}{\aap} \bibinfo{volume}{641}, \bibinfo{pages}{A166}.
\newblock \DOIprefix\doi{10.1051/0004-6361/202038689},
  \href{http://arxiv.org/abs/2006.10448}{{\tt arXiv:2006.10448}}.
\bibitem[{{Qiao} et~al.(2004){Qiao}, {Lee}, {Zhang}, {Xu} and
  {Wang}}]{2004ApJ...616L.127Q}
\bibinfo{author}{{Qiao}, G.J.}, \bibinfo{author}{{Lee}, K.J.},
  \bibinfo{author}{{Zhang}, B.}, \bibinfo{author}{{Xu}, R.X.},
  \bibinfo{author}{{Wang}, H.G.}, \bibinfo{year}{2004}.
\newblock \bibinfo{title}{{A Model for the Challenging ``Bi-drifting''
  Phenomenon in PSR J0815+09}}.
\newblock \bibinfo{journal}{\apjl} \bibinfo{volume}{616},
  \bibinfo{pages}{L127--L130}.
\newblock \DOIprefix\doi{10.1086/426862},
  \href{http://arxiv.org/abs/astro-ph/0410479}{{\tt arXiv:astro-ph/0410479}}.
\bibitem[{{Radhakrishnan} et~al.(1969){Radhakrishnan}, {Cooke}, {Komesaroff}
  and {Morris}}]{1969Natur.221..443R}
\bibinfo{author}{{Radhakrishnan}, V.}, \bibinfo{author}{{Cooke}, D.J.},
  \bibinfo{author}{{Komesaroff}, M.M.}, \bibinfo{author}{{Morris}, D.},
  \bibinfo{year}{1969}.
\newblock \bibinfo{title}{{Evidence in Support of a Rotational Model for the
  Pulsar PSR 0833-45}}.
\newblock \bibinfo{journal}{\nat} \bibinfo{volume}{221},
  \bibinfo{pages}{443--446}.
\newblock \DOIprefix\doi{10.1038/221443a0}.
\bibitem[{{Rankin}(1986)}]{1986ApJ...301..901R}
\bibinfo{author}{{Rankin}, J.M.}, \bibinfo{year}{1986}.
\newblock \bibinfo{title}{{Toward an Empirical Theory of Pulsar Emission. III.
  Mode Changing, Drifting Subpulses, and Pulse Nulling}}.
\newblock \bibinfo{journal}{\apj} \bibinfo{volume}{301}, \bibinfo{pages}{901}.
\newblock \DOIprefix\doi{10.1086/163955}.
\bibitem[{{Redman} et~al.(2005){Redman}, {Wright} and
  {Rankin}}]{2005MNRAS.357..859R}
\bibinfo{author}{{Redman}, S.L.}, \bibinfo{author}{{Wright}, G.A.E.},
  \bibinfo{author}{{Rankin}, J.M.}, \bibinfo{year}{2005}.
\newblock \bibinfo{title}{{Pulsar PSR B2303+30: a single system of drifting
  subpulses, moding and nulling}}.
\newblock \bibinfo{journal}{\mnras} \bibinfo{volume}{357},
  \bibinfo{pages}{859--872}.
\newblock \DOIprefix\doi{10.1111/j.1365-2966.2005.08672.x},
  \href{http://arxiv.org/abs/astro-ph/0407540}{{\tt arXiv:astro-ph/0407540}}.
\bibitem[{{Ruderman} and {Sutherland}(1975)}]{1975ApJ...196...51R}
\bibinfo{author}{{Ruderman}, M.A.}, \bibinfo{author}{{Sutherland}, P.G.},
  \bibinfo{year}{1975}.
\newblock \bibinfo{title}{{Theory of pulsars: polar gaps, sparks, and coherent
  microwave radiation.}}
\newblock \bibinfo{journal}{\apj} \bibinfo{volume}{196},
  \bibinfo{pages}{51--72}.
\newblock \DOIprefix\doi{10.1086/153393}.
\bibitem[{{Shannon}(1949)}]{1949IEEEP..37...10S}
\bibinfo{author}{{Shannon}, C.E.}, \bibinfo{year}{1949}.
\newblock \bibinfo{title}{{Communication in the Presence of Noise}}.
\newblock \bibinfo{journal}{IEEE Proceedings} \bibinfo{volume}{37},
  \bibinfo{pages}{10--21}.
\newblock \DOIprefix\doi{10.1109/JPROC.1998.659497}.
\bibitem[{{Song} et~al.(2023){Song}, {Weltevrede}, {Szary}, {Wright}, {Keith},
  {Basu}, {Johnston}, {Karastergiou}, {Main}, {Oswald}, {Parthasarathy},
  {Posselt}, {Bailes}, {Buchner}, {Hugo} and {Serylak}}]{2023MNRAS.520.4562S}
\bibinfo{author}{{Song}, X.}, \bibinfo{author}{{Weltevrede}, P.},
  \bibinfo{author}{{Szary}, A.}, \bibinfo{author}{{Wright}, G.},
  \bibinfo{author}{{Keith}, M.J.}, \bibinfo{author}{{Basu}, A.},
  \bibinfo{author}{{Johnston}, S.}, \bibinfo{author}{{Karastergiou}, A.},
  \bibinfo{author}{{Main}, R.A.}, \bibinfo{author}{{Oswald}, L.S.},
  \bibinfo{author}{{Parthasarathy}, A.}, \bibinfo{author}{{Posselt}, B.},
  \bibinfo{author}{{Bailes}, M.}, \bibinfo{author}{{Buchner}, S.},
  \bibinfo{author}{{Hugo}, B.}, \bibinfo{author}{{Serylak}, M.},
  \bibinfo{year}{2023}.
\newblock \bibinfo{title}{{The Thousand-Pulsar-Array programme on MeerKAT -
  VIII. The subpulse modulation of 1198 pulsars}}.
\newblock \bibinfo{journal}{\mnras} \bibinfo{volume}{520},
  \bibinfo{pages}{4562--4581}.
\newblock \DOIprefix\doi{10.1093/mnras/stad135},
  \href{http://arxiv.org/abs/2301.04067}{{\tt arXiv:2301.04067}}.
\bibitem[{{Stinebring} et~al.(2001){Stinebring}, {McLaughlin}, {Cordes},
  {Becker}, {Goodman}, {Kramer}, {Sheckard} and {Smith}}]{2001ApJ...549L..97S}
\bibinfo{author}{{Stinebring}, D.R.}, \bibinfo{author}{{McLaughlin}, M.A.},
  \bibinfo{author}{{Cordes}, J.M.}, \bibinfo{author}{{Becker}, K.M.},
  \bibinfo{author}{{Goodman}, J.E.E.}, \bibinfo{author}{{Kramer}, M.A.},
  \bibinfo{author}{{Sheckard}, J.L.}, \bibinfo{author}{{Smith}, C.T.},
  \bibinfo{year}{2001}.
\newblock \bibinfo{title}{{Faint Scattering Around Pulsars: Probing the
  Interstellar Medium on Solar System Size Scales}}.
\newblock \bibinfo{journal}{\apjl} \bibinfo{volume}{549},
  \bibinfo{pages}{L97--L100}.
\newblock \DOIprefix\doi{10.1086/319133},
  \href{http://arxiv.org/abs/astro-ph/0010363}{{\tt arXiv:astro-ph/0010363}}.
\bibitem[{{Sutton} et~al.(1970){Sutton}, {Staelin}, {Price} and
  {Weimer}}]{1970ApJ...159L..89S}
\bibinfo{author}{{Sutton}, J.M.}, \bibinfo{author}{{Staelin}, D.H.},
  \bibinfo{author}{{Price}, R.M.}, \bibinfo{author}{{Weimer}, R.},
  \bibinfo{year}{1970}.
\newblock \bibinfo{title}{{Three Pulsars with Marching Subpulses}}.
\newblock \bibinfo{journal}{\apjl} \bibinfo{volume}{159},
  \bibinfo{pages}{L89--L94}.
\newblock \DOIprefix\doi{10.1086/180484}.
\bibitem[{{Szary} and {van Leeuwen}(2017)}]{2017ApJ...845...95S}
\bibinfo{author}{{Szary}, A.}, \bibinfo{author}{{van Leeuwen}, J.},
  \bibinfo{year}{2017}.
\newblock \bibinfo{title}{{On the Origin of the Bi-drifting Subpulse Phenomenon
  in Pulsars}}.
\newblock \bibinfo{journal}{\apj} \bibinfo{volume}{845}, \bibinfo{pages}{95}.
\newblock \DOIprefix\doi{10.3847/1538-4357/aa803a},
  \href{http://arxiv.org/abs/1707.05046}{{\tt arXiv:1707.05046}}.
\bibitem[{{Taylor} and {Huguenin}(1971)}]{1971ApJ...167..273T}
\bibinfo{author}{{Taylor}, J.H.}, \bibinfo{author}{{Huguenin}, G.R.},
  \bibinfo{year}{1971}.
\newblock \bibinfo{title}{{Observations of Rapid Fluctuations of Intensity and
  Phase in Pulsar Emissions}}.
\newblock \bibinfo{journal}{\apj} \bibinfo{volume}{167}, \bibinfo{pages}{273}.
\newblock \DOIprefix\doi{10.1086/151030}.
\bibitem[{{Taylor} and {Jura}(1969)}]{1969Natur.223..797T}
\bibinfo{author}{{Taylor}, J.H.}, \bibinfo{author}{{Jura}, M.},
  \bibinfo{year}{1969}.
\newblock \bibinfo{title}{{Periodic Intensity Fluctuations in Pulsars}}.
\newblock \bibinfo{journal}{\nat} \bibinfo{volume}{223},
  \bibinfo{pages}{797--799}.
\newblock \DOIprefix\doi{10.1038/223797a0}.
\bibitem[{{Tian} et~al.(2023a){Tian}, {Zhang}, {Wang}, {Wang}, {Sun}, {Liu},
  {Zhang}, {Dai}, {Yuan}, {Zhang}, {Liu}, {Jiang}, {Wu}, {Zheng}, {Chen}, {Li},
  {Zhu}, {Pan}, {Gan}, {Chen} and {Sai}}]{2023arXiv230714015T}
\bibinfo{author}{{Tian}, P.}, \bibinfo{author}{{Zhang}, P.},
  \bibinfo{author}{{Wang}, W.}, \bibinfo{author}{{Wang}, P.},
  \bibinfo{author}{{Sun}, X.}, \bibinfo{author}{{Liu}, J.},
  \bibinfo{author}{{Zhang}, B.}, \bibinfo{author}{{Dai}, Z.},
  \bibinfo{author}{{Yuan}, F.}, \bibinfo{author}{{Zhang}, S.},
  \bibinfo{author}{{Liu}, Q.}, \bibinfo{author}{{Jiang}, P.},
  \bibinfo{author}{{Wu}, X.}, \bibinfo{author}{{Zheng}, Z.},
  \bibinfo{author}{{Chen}, J.}, \bibinfo{author}{{Li}, D.},
  \bibinfo{author}{{Zhu}, Z.}, \bibinfo{author}{{Pan}, Z.},
  \bibinfo{author}{{Gan}, H.}, \bibinfo{author}{{Chen}, X.},
  \bibinfo{author}{{Sai}, N.}, \bibinfo{year}{2023}a.
\newblock \bibinfo{title}{{Sub-second periodic radio oscillations in a
  microquasar}}.
\newblock \bibinfo{journal}{Nature} \bibinfo{volume}{621},
  \bibinfo{pages}{271--275}.
\newblock \DOIprefix\doi{10.48550/arXiv.2307.14015},
  \href{http://arxiv.org/abs/2307.14015}{{\tt arXiv:2307.14015}}.
\bibitem[{{Tian} et~al.(2023b){Tian}, {Zhang}, {Wang}, {Wang}, {Sun} and
  {Zheng}}]{2023JHEAp..39...43T}
\bibinfo{author}{{Tian}, P.F.}, \bibinfo{author}{{Zhang}, P.},
  \bibinfo{author}{{Wang}, W.}, \bibinfo{author}{{Wang}, P.},
  \bibinfo{author}{{Sun}, X.}, \bibinfo{author}{{Zheng}, Z.},
  \bibinfo{year}{2023}b.
\newblock \bibinfo{title}{{The radio monitoring of the microquasar GRS 1915+105
  with FAST}}.
\newblock \bibinfo{journal}{Journal of High Energy Astrophysics}
  \bibinfo{volume}{39}, \bibinfo{pages}{43--52}.
\newblock \DOIprefix\doi{10.1016/j.jheap.2023.06.002}.
\bibitem[{{Torrence} and {Compo}(1998)}]{1998BAMS...79...61T}
\bibinfo{author}{{Torrence}, C.}, \bibinfo{author}{{Compo}, G.P.},
  \bibinfo{year}{1998}.
\newblock \bibinfo{title}{{A Practical Guide to Wavelet Analysis.}}
\newblock \bibinfo{journal}{Bulletin of the American Meteorological Society}
  \bibinfo{volume}{79}, \bibinfo{pages}{61--78}.
\newblock \DOIprefix\doi{10.1175/1520-0477(1998)079<0061:APGTWA>2.0.CO;2}.
\bibitem[{{van Leeuwen} and {Timokhin}(2012)}]{2012ApJ...752..155V}
\bibinfo{author}{{van Leeuwen}, J.}, \bibinfo{author}{{Timokhin}, A.N.},
  \bibinfo{year}{2012}.
\newblock \bibinfo{title}{{On Plasma Rotation and Drifting Subpulses in
  Pulsars: Using Aligned Pulsar B0826-34 as a Voltmeter}}.
\newblock \bibinfo{journal}{\apj} \bibinfo{volume}{752}, \bibinfo{pages}{155}.
\newblock \DOIprefix\doi{10.1088/0004-637X/752/2/155},
  \href{http://arxiv.org/abs/1201.3647}{{\tt arXiv:1201.3647}}.
\bibitem[{{van Straten} and {Bailes}(2011)}]{2011PASA...28....1V}
\bibinfo{author}{{van Straten}, W.}, \bibinfo{author}{{Bailes}, M.},
  \bibinfo{year}{2011}.
\newblock \bibinfo{title}{{DSPSR: Digital Signal Processing Software for Pulsar
  Astronomy}}.
\newblock \bibinfo{journal}{\pasa} \bibinfo{volume}{28},
  \bibinfo{pages}{1--14}.
\newblock \DOIprefix\doi{10.1071/AS10021},
  \href{http://arxiv.org/abs/1008.3973}{{\tt arXiv:1008.3973}}.
\bibitem[{{Wang} et~al.(2005){Wang}, {Manchester}, {Johnston}, {Rickett},
  {Zhang}, {Yusup} and {Chen}}]{2005MNRAS.358..270W}
\bibinfo{author}{{Wang}, N.}, \bibinfo{author}{{Manchester}, R.N.},
  \bibinfo{author}{{Johnston}, S.}, \bibinfo{author}{{Rickett}, B.},
  \bibinfo{author}{{Zhang}, J.}, \bibinfo{author}{{Yusup}, A.},
  \bibinfo{author}{{Chen}, M.}, \bibinfo{year}{2005}.
\newblock \bibinfo{title}{{Long-term scintillation observations of five pulsars
  at 1540 MHz}}.
\newblock \bibinfo{journal}{\mnras} \bibinfo{volume}{358},
  \bibinfo{pages}{270--282}.
\newblock \DOIprefix\doi{10.1111/j.1365-2966.2005.08798.x},
  \href{http://arxiv.org/abs/astro-ph/0501218}{{\tt arXiv:astro-ph/0501218}}.
\bibitem[{{Weltevrede} et~al.(2006){Weltevrede}, {Edwards} and
  {Stappers}}]{2006A&A...445..243W}
\bibinfo{author}{{Weltevrede}, P.}, \bibinfo{author}{{Edwards}, R.T.},
  \bibinfo{author}{{Stappers}, B.W.}, \bibinfo{year}{2006}.
\newblock \bibinfo{title}{{The subpulse modulation properties of pulsars at 21
  cm}}.
\newblock \bibinfo{journal}{\aap} \bibinfo{volume}{445},
  \bibinfo{pages}{243--272}.
\newblock \DOIprefix\doi{10.1051/0004-6361:20053088},
  \href{http://arxiv.org/abs/astro-ph/0507282}{{\tt arXiv:astro-ph/0507282}}.
\bibitem[{{Weltevrede} et~al.(2011){Weltevrede}, {Johnston} and
  {Espinoza}}]{2011MNRAS.411.1917W}
\bibinfo{author}{{Weltevrede}, P.}, \bibinfo{author}{{Johnston}, S.},
  \bibinfo{author}{{Espinoza}, C.M.}, \bibinfo{year}{2011}.
\newblock \bibinfo{title}{{The glitch-induced identity changes of PSR
  J1119-6127}}.
\newblock \bibinfo{journal}{\mnras} \bibinfo{volume}{411},
  \bibinfo{pages}{1917--1934}.
\newblock \DOIprefix\doi{10.1111/j.1365-2966.2010.17821.x},
  \href{http://arxiv.org/abs/1010.0857}{{\tt arXiv:1010.0857}}.
\bibitem[{{Weltevrede} et~al.(2007){Weltevrede}, {Stappers} and
  {Edwards}}]{2007A&A...469..607W}
\bibinfo{author}{{Weltevrede}, P.}, \bibinfo{author}{{Stappers}, B.W.},
  \bibinfo{author}{{Edwards}, R.T.}, \bibinfo{year}{2007}.
\newblock \bibinfo{title}{{The subpulse modulation properties of pulsars at 92
  cm and the frequency dependence of subpulse modulation}}.
\newblock \bibinfo{journal}{\aap} \bibinfo{volume}{469},
  \bibinfo{pages}{607--631}.
\newblock \DOIprefix\doi{10.1051/0004-6361:20066855},
  \href{http://arxiv.org/abs/0704.3572}{{\tt arXiv:0704.3572}}.
\bibitem[{{Wright}(2022)}]{2022MNRAS.514.4046W}
\bibinfo{author}{{Wright}, G.}, \bibinfo{year}{2022}.
\newblock \bibinfo{title}{{Pulsar emission patterns seen as evidence for
  magnetospheric interactions}}.
\newblock \bibinfo{journal}{\mnras} \bibinfo{volume}{514},
  \bibinfo{pages}{4046--4060}.
\newblock \DOIprefix\doi{10.1093/mnras/stac1629}.
\bibitem[{{Wright} and {Weltevrede}(2017)}]{2017MNRAS.464.2597W}
\bibinfo{author}{{Wright}, G.}, \bibinfo{author}{{Weltevrede}, P.},
  \bibinfo{year}{2017}.
\newblock \bibinfo{title}{{Pulsar bi-drifting: implications for polar cap
  geometry}}.
\newblock \bibinfo{journal}{\mnras} \bibinfo{volume}{464},
  \bibinfo{pages}{2597--2608}.
\newblock \DOIprefix\doi{10.1093/mnras/stw2498},
  \href{http://arxiv.org/abs/1609.09241}{{\tt arXiv:1609.09241}}.
\bibitem[{{Xu} et~al.(2022){Xu}, {Niu}, {Chen}, {Lee}, {Zhu}, {Dong}, {Zhang},
  {Jiang}, {Wang}, {Xu}, {Zhang}, {Fu}, {Filippenko}, {Peng}, {Zhou}, {Zhang},
  {Wang}, {Feng}, {Li}, {Brink}, {Li}, {Lu}, {Yang}, {Caballero}, {Cai},
  {Chen}, {Dai}, {Djorgovski}, {Esamdin}, {Gan}, {Guhathakurta}, {Han}, {Hao},
  {Huang}, {Jiang}, {Li}, {Li}, {Li}, {Li}, {Li}, {Liu}, {Luo}, {Men}, {Niu},
  {Peng}, {Qian}, {Song}, {Stern}, {Stockton}, {Sun}, {Wang}, {Wang}, {Wang},
  {Wang}, {Wu}, {Xiao}, {Xiong}, {Xu}, {Xu}, {Yang}, {Yang}, {Yao}, {Yi},
  {Yue}, {Yu}, {Yu}, {Yuan}, {Zhang}, {Zhang}, {Zhang}, {Zhao}, {Zheng}, {Zhu}
  and {Zou}}]{2022Natur.609..685X}
\bibinfo{author}{{Xu}, H.}, \bibinfo{author}{{Niu}, J.R.},
  \bibinfo{author}{{Chen}, P.}, \bibinfo{author}{{Lee}, K.J.},
  \bibinfo{author}{{Zhu}, W.W.}, \bibinfo{author}{{Dong}, S.},
  \bibinfo{author}{{Zhang}, B.}, \bibinfo{author}{{Jiang}, J.C.},
  \bibinfo{author}{{Wang}, B.J.}, \bibinfo{author}{{Xu}, J.W.},
  \bibinfo{author}{{Zhang}, C.F.}, \bibinfo{author}{{Fu}, H.},
  \bibinfo{author}{{Filippenko}, A.V.}, \bibinfo{author}{{Peng}, E.W.},
  \bibinfo{author}{{Zhou}, D.J.}, \bibinfo{author}{{Zhang}, Y.K.},
  \bibinfo{author}{{Wang}, P.}, \bibinfo{author}{{Feng}, Y.},
  \bibinfo{author}{{Li}, Y.}, \bibinfo{author}{{Brink}, T.G.},
  \bibinfo{author}{{Li}, D.Z.}, \bibinfo{author}{{Lu}, W.},
  \bibinfo{author}{{Yang}, Y.P.}, \bibinfo{author}{{Caballero}, R.N.},
  \bibinfo{author}{{Cai}, C.}, \bibinfo{author}{{Chen}, M.Z.},
  \bibinfo{author}{{Dai}, Z.G.}, \bibinfo{author}{{Djorgovski}, S.G.},
  \bibinfo{author}{{Esamdin}, A.}, \bibinfo{author}{{Gan}, H.Q.},
  \bibinfo{author}{{Guhathakurta}, P.}, \bibinfo{author}{{Han}, J.L.},
  \bibinfo{author}{{Hao}, L.F.}, \bibinfo{author}{{Huang}, Y.X.},
  \bibinfo{author}{{Jiang}, P.}, \bibinfo{author}{{Li}, C.K.},
  \bibinfo{author}{{Li}, D.}, \bibinfo{author}{{Li}, H.},
  \bibinfo{author}{{Li}, X.Q.}, \bibinfo{author}{{Li}, Z.X.},
  \bibinfo{author}{{Liu}, Z.Y.}, \bibinfo{author}{{Luo}, R.},
  \bibinfo{author}{{Men}, Y.P.}, \bibinfo{author}{{Niu}, C.H.},
  \bibinfo{author}{{Peng}, W.X.}, \bibinfo{author}{{Qian}, L.},
  \bibinfo{author}{{Song}, L.M.}, \bibinfo{author}{{Stern}, D.},
  \bibinfo{author}{{Stockton}, A.}, \bibinfo{author}{{Sun}, J.H.},
  \bibinfo{author}{{Wang}, F.Y.}, \bibinfo{author}{{Wang}, M.},
  \bibinfo{author}{{Wang}, N.}, \bibinfo{author}{{Wang}, W.Y.},
  \bibinfo{author}{{Wu}, X.F.}, \bibinfo{author}{{Xiao}, S.},
  \bibinfo{author}{{Xiong}, S.L.}, \bibinfo{author}{{Xu}, Y.H.},
  \bibinfo{author}{{Xu}, R.X.}, \bibinfo{author}{{Yang}, J.},
  \bibinfo{author}{{Yang}, X.}, \bibinfo{author}{{Yao}, R.},
  \bibinfo{author}{{Yi}, Q.B.}, \bibinfo{author}{{Yue}, Y.L.},
  \bibinfo{author}{{Yu}, D.J.}, \bibinfo{author}{{Yu}, W.F.},
  \bibinfo{author}{{Yuan}, J.P.}, \bibinfo{author}{{Zhang}, B.B.},
  \bibinfo{author}{{Zhang}, S.B.}, \bibinfo{author}{{Zhang}, S.N.},
  \bibinfo{author}{{Zhao}, Y.}, \bibinfo{author}{{Zheng}, W.K.},
  \bibinfo{author}{{Zhu}, Y.}, \bibinfo{author}{{Zou}, J.H.},
  \bibinfo{year}{2022}.
\newblock \bibinfo{title}{{A fast radio burst source at a complex magnetized
  site in a barred galaxy}}.
\newblock \bibinfo{journal}{\nat} \bibinfo{volume}{609},
  \bibinfo{pages}{685--688}.
\newblock \DOIprefix\doi{10.1038/s41586-022-05071-8},
  \href{http://arxiv.org/abs/2111.11764}{{\tt arXiv:2111.11764}}.
\bibitem[{{Zhang} et~al.(2019){Zhang}, {Li}, {Hobbs}, {Agar}, {Manchester},
  {Weltevrede}, {Coles}, {Wang}, {Zhu}, {Wen}, {Yuan}, {Cameron}, {Dai}, {Liu},
  {Zhi}, {Miao}, {Yuan}, {Cao}, {Feng}, {Gan}, {Gao}, {Gu}, {Guo}, {Hao},
  {Huang}, {Jiang}, {Jin}, {Li}, {Li}, {Li}, {Liu}, {Pan}, {Pan}, {Peng},
  {Qian}, {Qian}, {Shi}, {Song}, {Song}, {Sun}, {Sun}, {Wang}, {Wang}, {Wang},
  {Xie}, {Yan}, {Yang}, {Yang}, {Yao}, {Yu}, {Yu}, {Yue}, {Zhang}, {Zhang},
  {Zhang}, {Zheng}, {Zhou}, {Zhu}, {Zhu}, {Zhu}, {Zhu} and
  {Zhu}}]{2019ApJ...877...55Z}
\bibinfo{author}{{Zhang}, L.}, \bibinfo{author}{{Li}, D.},
  \bibinfo{author}{{Hobbs}, G.}, \bibinfo{author}{{Agar}, C.H.},
  \bibinfo{author}{{Manchester}, R.N.}, \bibinfo{author}{{Weltevrede}, P.},
  \bibinfo{author}{{Coles}, W.A.}, \bibinfo{author}{{Wang}, P.},
  \bibinfo{author}{{Zhu}, W.}, \bibinfo{author}{{Wen}, Z.},
  \bibinfo{author}{{Yuan}, J.}, \bibinfo{author}{{Cameron}, A.D.},
  \bibinfo{author}{{Dai}, S.}, \bibinfo{author}{{Liu}, K.},
  \bibinfo{author}{{Zhi}, Q.}, \bibinfo{author}{{Miao}, C.},
  \bibinfo{author}{{Yuan}, M.}, \bibinfo{author}{{Cao}, S.},
  \bibinfo{author}{{Feng}, L.}, \bibinfo{author}{{Gan}, H.},
  \bibinfo{author}{{Gao}, L.}, \bibinfo{author}{{Gu}, X.},
  \bibinfo{author}{{Guo}, M.}, \bibinfo{author}{{Hao}, Q.},
  \bibinfo{author}{{Huang}, L.}, \bibinfo{author}{{Jiang}, P.},
  \bibinfo{author}{{Jin}, C.}, \bibinfo{author}{{Li}, H.},
  \bibinfo{author}{{Li}, Q.}, \bibinfo{author}{{Li}, Q.},
  \bibinfo{author}{{Liu}, H.}, \bibinfo{author}{{Pan}, G.},
  \bibinfo{author}{{Pan}, Z.}, \bibinfo{author}{{Peng}, B.},
  \bibinfo{author}{{Qian}, H.}, \bibinfo{author}{{Qian}, L.},
  \bibinfo{author}{{Shi}, X.}, \bibinfo{author}{{Song}, J.},
  \bibinfo{author}{{Song}, L.}, \bibinfo{author}{{Sun}, C.},
  \bibinfo{author}{{Sun}, J.}, \bibinfo{author}{{Wang}, H.},
  \bibinfo{author}{{Wang}, Q.}, \bibinfo{author}{{Wang}, Y.},
  \bibinfo{author}{{Xie}, X.}, \bibinfo{author}{{Yan}, J.},
  \bibinfo{author}{{Yang}, L.}, \bibinfo{author}{{Yang}, S.},
  \bibinfo{author}{{Yao}, R.}, \bibinfo{author}{{Yu}, D.},
  \bibinfo{author}{{Yu}, J.}, \bibinfo{author}{{Yue}, Y.},
  \bibinfo{author}{{Zhang}, C.}, \bibinfo{author}{{Zhang}, H.},
  \bibinfo{author}{{Zhang}, S.}, \bibinfo{author}{{Zheng}, X.},
  \bibinfo{author}{{Zhou}, A.}, \bibinfo{author}{{Zhu}, B.},
  \bibinfo{author}{{Zhu}, L.}, \bibinfo{author}{{Zhu}, M.},
  \bibinfo{author}{{Zhu}, W.}, \bibinfo{author}{{Zhu}, Y.},
  \bibinfo{year}{2019}.
\newblock \bibinfo{title}{{PSR J1926-0652: A Pulsar with Interesting Emission
  Properties Discovered at FAST}}.
\newblock \bibinfo{journal}{\apj} \bibinfo{volume}{877}, \bibinfo{pages}{55}.
\newblock \DOIprefix\doi{10.3847/1538-4357/ab1849},
  \href{http://arxiv.org/abs/1904.05482}{{\tt arXiv:1904.05482}}.

\end{thebibliography}

\bio{}
\endbio

\bio{}
\endbio

\end{document}